\documentclass[preprint]{aastex62}

\usepackage{amssymb}
\usepackage[caption=false]{subfig} 
\usepackage{float}
\usepackage{gensymb}

\shorttitle{3.8~$\mu$m Imaging of 400 -- 600~K Brown Dwarfs}
\shortauthors{Leggett et al.}

\begin{document}


\title{3.8~$\mu$m Imaging of 400 -- 600~K Brown Dwarfs and\\Orbital Constraints for WISEP J045853.90+643452.6AB}


\author{S. K. Leggett}
\affiliation{Gemini Observatory, Northern Operations Center,  670 N. A'ohoku Place, Hilo, HI 96720, USA}
\author{Trent J. Dupuy}
\affiliation{Gemini Observatory, Northern Operations Center,  670 N. A'ohoku Place, Hilo, HI 96720, USA}
\author{Caroline V. Morley}
\affiliation{University of Texas at Austin, Austin, TX 78712, USA}
\author{Mark S. Marley}
\affiliation{NASA Ames Research Center, Mail Stop 245-3, Moffett Field, CA 94035, USA }
\author{William M. J. Best}
\affiliation{University of Texas at Austin, Austin, TX 78712, USA}
\author{Michael C. Liu}
\affiliation{Institute for Astronomy, University of Hawaii at Manoa, Honolulu, HI 96822, USA}
\author{D. Apai}
\affiliation{Department of Astronomy/Steward Observatory, University of Arizona, 933 N. Cherry Avenue, Tucson, AZ 85721, USA ; Department of Planetary Science/Lunar and Planetary Laboratory, University of Arizona, 1640 E. University Boulevard, Tucson, AZ 85718, USA ; Earths in Other Solar Systems Team, NASA Nexus for Exoplanet System Science, USA 0000-0003-3714-5855}
\author{S. L. Casewell}
\affiliation{Department of Physics and Astronomy, University of Leicester, University Road, Leicester LE1 7RH, UK}
\author{T. R. Geballe}
\affiliation{Gemini Observatory, Northern Operations Center,  670 N. A'ohoku Place, Hilo, HI 96720, USA}
\author{John E. Gizis}
\affiliation{Department of Physics and Astronomy, University of Delaware, Newark, DE 19716, USA}
\author{J. Sebastian Pineda}
\affiliation{University of Colorado Boulder, Laboratory for Atmospheric and Space Physics, 3665 Discovery Drive, Boulder, CO 80303, USA}
\author{Marcia Rieke}
\affiliation{Steward Observatory, University of Arizona, 933 N. Cherry Avenue, Tucson, AZ 85721, USA}
\author{G. S. Wright}
\affiliation{STFC UK-ATC, Edinburgh, EH9 3HJ, UK}

\begin{abstract}

Half of the energy emitted by late-T- and Y-type brown dwarfs emerges at $3.5 \leq \lambda~\mu$m $\leq 5.5$. 
We present new  $L^{\prime}$ ($3.43 \leq \lambda~\mu$m$~\leq 4.11$) photometry obtained at the Gemini North 
telescope for nine late-T and Y dwarfs, and synthesize $L^{\prime}$ from spectra for an additional two dwarfs. The 
targets include two binary systems which were imaged at a resolution of 0$\farcs$25. One of these,  
WISEP J045853.90$+$643452.6AB, shows significant motion, and we present an astrometric analysis of the binary
using {\it Hubble Space Telescope}, Keck Adaptive Optics, and Gemini images.  We compare $\lambda\sim 4~\mu$m 
observations to models, and find that the model fluxes are too low for brown dwarfs cooler than $\sim$700~K.
The discrepancy increases with decreasing temperature, and is  a factor of 
$\sim$2 at $T_{\rm eff} = 500$~K and $\sim$4 at $T_{\rm eff} = 400$~K.  Warming the upper layers of a model 
atmosphere generates a spectrum closer to what is observed. The thermal structure of cool brown dwarf atmospheres 
above the radiative-convective boundary may not be adequately modelled using pure radiative equilibrium; instead 
heat may be introduced by thermochemical instabilities (previously suggested for the L- to T-type transition) or 
by breaking gravity waves (previously suggested for the solar system giant planets). One-dimensional models may 
not capture these atmospheres, which likely have both horizontal and vertical pressure/temperature variations. 

\end{abstract}

\section{Introduction}

Brown dwarfs form the extended low-mass tail of the stellar initial mass function, and brown dwarfs as low-mass as four Jupiter-masses ($M_{\rm Jup}$) have been found in young clusters and associations \citep[e.g.,][]{Best2017,Esplin2017}. The difference between giant planet and brown dwarf formation is an active research area \citep[e.g.,][]{Nielsen2019,Schlaufman2018,Wagner2019}.  Brown dwarfs have the compositions of stars, but the physics and chemistry of their atmospheres are complex and resemble those of giant planets \citep[e.g.,][]{Line2015,Morley2014a}.

\begin{figure}[!b]
\begin{center}
\vskip -1.3in
\hskip -0.5in
\includegraphics[angle=-90,width=1.05\textwidth]{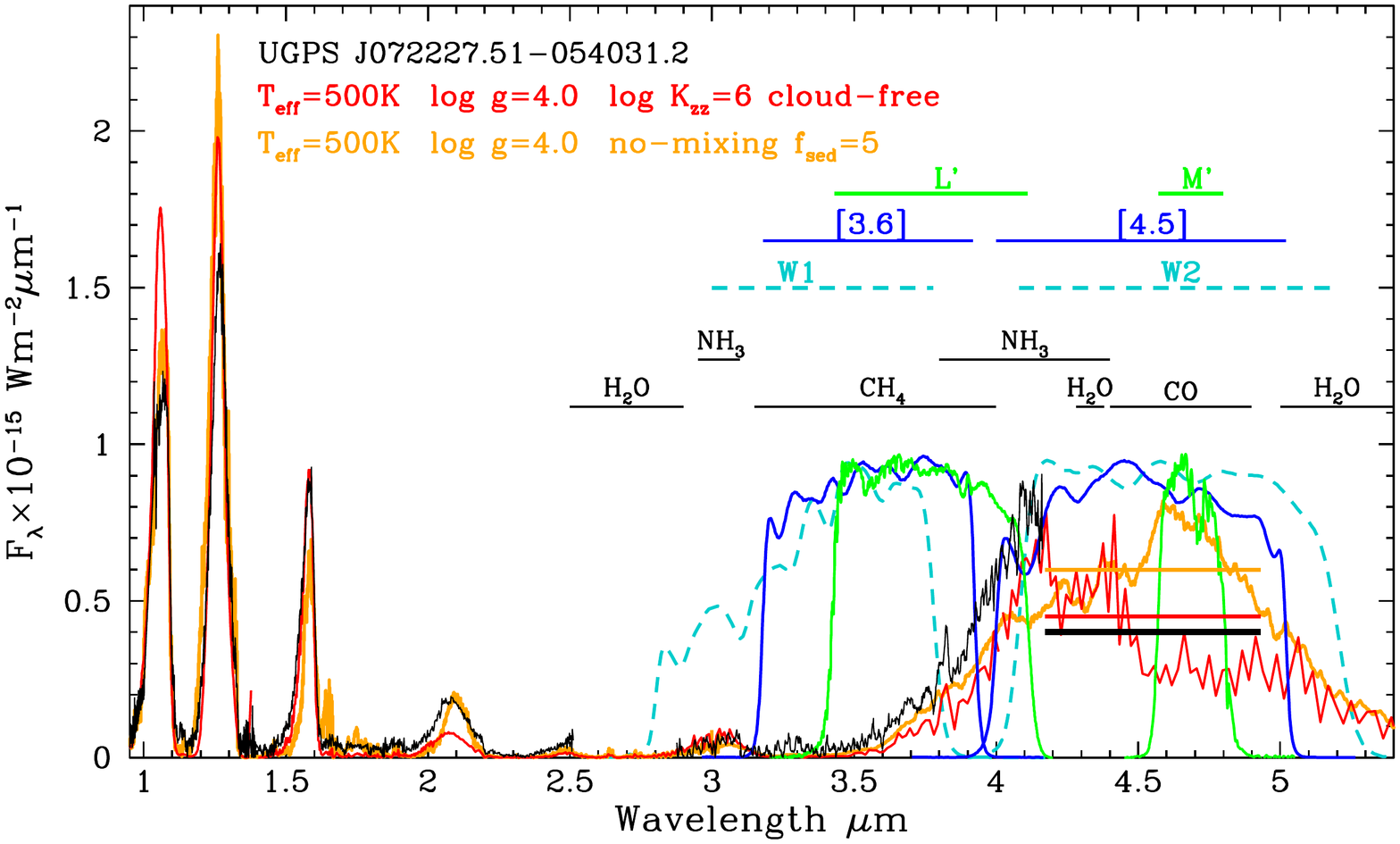}
\vskip -0.25in
\caption{Comparison of the observed spectrum of the 500~K brown dwarf UGPS J072227.51$-$054031.2 \citep[black line,][]{Leggett2012} with model spectra by
\citet[][red line]{Tremblin2015} and \citet[][orange line]{Morley2012}. The horizontal lines at $4 \lesssim \lambda~\mu$m $\lesssim 5$
indicate the calculated and observed [4.5] magnitudes; the width of the black bar shows the uncertainty in the observed flux.  Filter bandpasses at $ 3 \lesssim \lambda~\mu$m $\lesssim 5$ are shown:
{\it WISE} W1 and W2 (cyan), {\it Spitzer} [3.6] and [4.5] (blue), and MKO $L^{\prime}$ and $M^{\prime}$ (green). The primary molecular absorbers for this region are identified.
The model spectrum has been scaled for 
the measured distance of the brown dwarf and the evolutionary-implied radius that corresponds to the model temperature and gravity. 
Discrepancies between the models and observations are discussed in Section 5.
}
\end{center}
\end{figure}

The thermal radiation of a brown dwarf escapes through an atmosphere that is rich in molecules. Light emerges through windows between the broad molecular absorption bands. One of these windows spans $3.5 \leq \lambda~\mu$m $\leq 5.5$; for late-type T and Y dwarfs with effective temperature ($T_{\rm eff}$) less than 600~K \citep[e.g.,][]{Beichman2014,Dupuy2013,Kirkpatrick2012}, 40 -- 60\% of the total energy emerges through this single window.  The W2 filter of the
{\it Wide-field Infrared Survey Explorer} \citep[{\it WISE},][]{Wright2010} is centered at $\lambda \sim 4.6~\mu$m and the coldest known objects outside of the solar system have been discovered by  {\it WISE}  \citep{Cushing2011,Luhman2014}.
This paper presents new ground-based photometry of cool brown dwarfs using the Maunakea Observatories (MKO) $L^{\prime}$ filter, which is centered at $\lambda \sim 3.8~\mu$m \citep{Tokunaga2002}.  In this paper we combine the new photometry with published data  to explore the important $\lambda \sim 4~\mu$m spectral region.

Figure 1 shows the observed and synthetic $0.95 \leq \lambda~\mu$m $\leq 5.40$ spectrum of the 500~K brown dwarf UGPS J072227.51$-$054031.2 \citep[hereafter 0722,][]{Leggett2012,Lucas2010}. The bandpasses for the {\it Spitzer} [3.6] and [4.5], {\it WISE} W1 and W2, and ground-based MKO $L^{\prime}$ 
and $M^{\prime}$ filters are also shown. The figure  shows that these bandpasses sample slightly different regions of the $4~\mu$m spectrum and shows also that the spectrum is heavily sculpted by strong absorption bands. At $\lambda \approx 3.3~\mu$m the atmosphere is opaque; this is the region with the strong P-, Q- and R-branches of the $\nu_3$ band of methane and the flux emerges from the cold upper layers of the atmosphere. At  $\lambda \approx 4.1~\mu$m the brown dwarf is bright, the atmosphere is transparent, and flux emerges from very deep, hot, high-pressure regions of the atmosphere. The 3 -- 5~$\mu$m colors of brown dwarfs can therefore provide information on very physically different layers of the atmosphere.
The figure also illustrates the long-standing problem,  for brown dwarfs cooler than 700~K,
of significant shortfall in modelled flux at 
$3.4 \lesssim \lambda~\mu$m $\lesssim 4.1$ \citep{Leggett2010a,Leggett2012,Leggett2013,Leggett2015,Leggett2017,Luhman2016}. We explore this problem later in the paper.

Although space missions are more sensitive than ground-based observatories in the mid-infrared, the  {\it WISE} and {\it Spitzer}  cameras have relatively poor angular resolution with full-width half-maxima (FWHMs) of $\sim 6\farcs0$ for W1 and W2 \citep{Wright2010} and  $\sim 1\farcs7$ for [3.6] and [4.5] \citep{Fazio2004}.
This means that source confusion can be a problem, and close binary systems cannot be resolved. Cold brown dwarfs and in particular binaries containing cold brown dwarfs, which can test both atmospheric and evolutionary models, are prime targets for the upcoming {\it James Webb Space Telescope} ({\it JWST}) mission. The brown dwarf imaging presented here has a resolution of 0$\farcs$25 to 1$\farcs$0 and
will be used to improve {\it JWST} observation specifications.

Section 2 of this paper presents the new $L^{\prime}$ photometry. Section 3 gives the astrometric data for 
WISEP J045853.90+643452.6AB  (hereafter 0458AB) and presents preliminary orbital constraints for the system. In Section 4 we use astrometric and photometric properties to constrain the masses and ages of the 0458AB and
WISEPC J121756.91$+$162640.2AB  (hereafter 1217AB) binary systems. In Section 5 we examine the discrepancy between the models and the observations in this waveband. Our conclusions are given in Section 6. The Appendix presents ground-based and space-based 3 -- 5~$\mu$m colors for T and Y dwarfs, so that these datasets can be more easily utilized in the future.

\section{$L^{\prime}$ Photometry of Brown Dwarfs}

\subsection{Existing Data}

$L^{\prime}$ photometry of brown dwarfs with spectral type later than T6 has been published by: \citet{Burningham2009,Geballe2001,Golimowski2004,Leggett2002,Leggett2007}. In the next two subsections we present new observations and synthesized $L^{\prime}$ photometry from observed spectra. Figure 2 is a color-magnitude diagram which shows $M_{[4.5]}$ as a function of [3.6] $-$ [4.5] (see Figure 1 for bandpasses). The {\it Spitzer} data 
are taken from  \citet{Kirkpatrick2019,Leggett2017,Martin2018} and references therein, the trigonometric parallaxes are taken from \citet{Kirkpatrick2019,Leggett2017,Martin2018,Smart2018,Theissen2018} and references therein. T and Y dwarfs with $L^{\prime}$ photometry presented here are identified.

\begin{figure}[!b]
\begin{center}
\vskip -0.2in
\includegraphics[angle=0,width=0.7\textwidth]{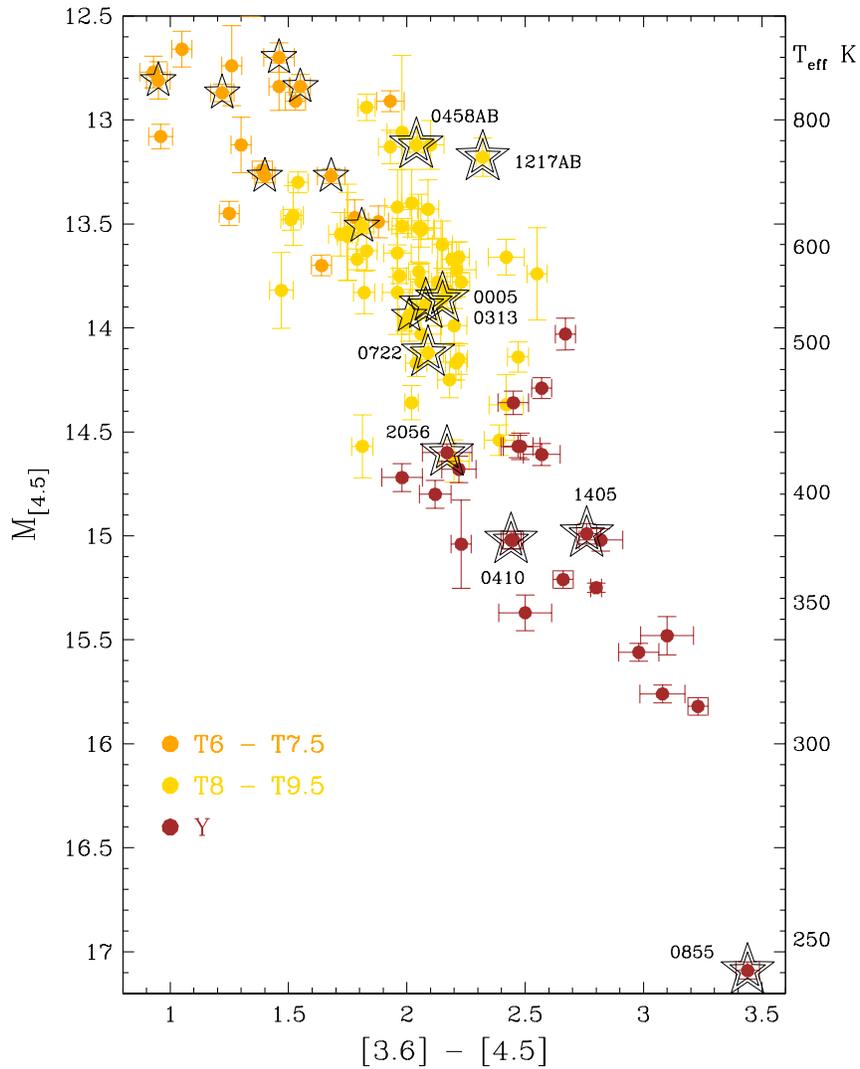}
\vskip -0.45in
\caption{$M_{[4.5]}$ as a function of [3.6] $-$ [4.5] for T and Y dwarfs. Starred symbols have measured  $L^{\prime}$; double stars represent targets with new data presented here, which are identified by the first four digits of their RA. $T_{\rm eff}$ values on the right axis are from \citet{Tremblin2015} non-equilibrium chemistry models. 
}
\end{center}
\end{figure}

\begin{deluxetable}{llcr}[!t]
\tablewidth{0pt}
\tablecaption{Observation Log}
\tablehead{
\colhead{\it WISE} & \colhead{Program} & \colhead{Date} & \colhead{Exposure} \\
\colhead{Name} & \colhead{Number} & \colhead{YYYMMDD} & \colhead{minutes} 
}
\startdata 
J000517.48$+$373720.5 & GN-2018B-FT-112 & 20181207 & 8.9 \\ 
J031325.96$+$780744.2 & GN-2018B-FT-112 & 20181223 & 9.2 \\ 
J041022.75$+$150247.9 & GN-2018B-FT-112 & 20181207 & 39.0 \\ 
J045853.90+643452.6AB & GN-2017B-FT-15 & 20171224 & 22.8 \\ 
J121756.91$+$162640.2AB & GN-2017B-FT-15 & 20180102 & 26.0 \\ 
J140518.32$+$553421.3 & GN-2018B-FT-112 & 20181223 & 31.4 \\ 
J205628.88$+$145953.6& GN-2018B-FT-112 & 20181207 & 15.2 
\enddata 
\end{deluxetable}

\subsection{Gemini Observations}

$L^{\prime}$ photometry was obtained using the Gemini Observatory  near-infrared imager  \citep[NIRI,][]{Hodapp2003} on the Gemini North telescope. Two binary systems that are targets for {\it JWST} Guaranteed Time Observations were observed via program GN-2017B-FT-15 in excellent natural seeing in order to resolve the binary components: 0458AB  and 1217AB.  Five (notionally) single brown dwarfs were observed via program GN-2018B-FT-112 in poorer seeing:
WISE J000517.48$+$373720.5, WISEPA J031325.96$+$780744.2, WISEA J041022.75$+$150247.9, WISEA J140518.32$+$553421.3 and WISEA J205628.88$+$145953.6. In the rest of this paper we shorten the object names to the first four digits of the Right Ascension values.  For GN-2018B-FT-112, brown dwarfs later than T8 with [3.6] $<$ 17 mag
that were accessible at the Gemini North telescope were selected. The targets were also chosen in order to sample the brightness and color space of Figure 2. No evidence of binarity was seen in the GN-2018B-FT-112 images, for which the seeing was typically  $1\farcs0$.

Table 1 lists the targets, program number, date of observation, and total exposure time. All nights were photometric. Individual exposures were 19~s, composed of 24 coadded 0.8~s frames. NIRI was used in f/32 mode,  with a pixel size of $0\farcs02$ and a field of view of 20''.  Targets and photometric standards  
were observed in a fixed 5-position grid pattern with 3'' telescope offsets. The photometric standards were selected from the \citet{Leggett2003} catalog
and were observed immediately before or after the brown dwarf at a similar airmass.

\begin{figure}[!b]
\vskip -0.7in
\hskip 0.1in
    \includegraphics[angle=0,width=0.48\textwidth]{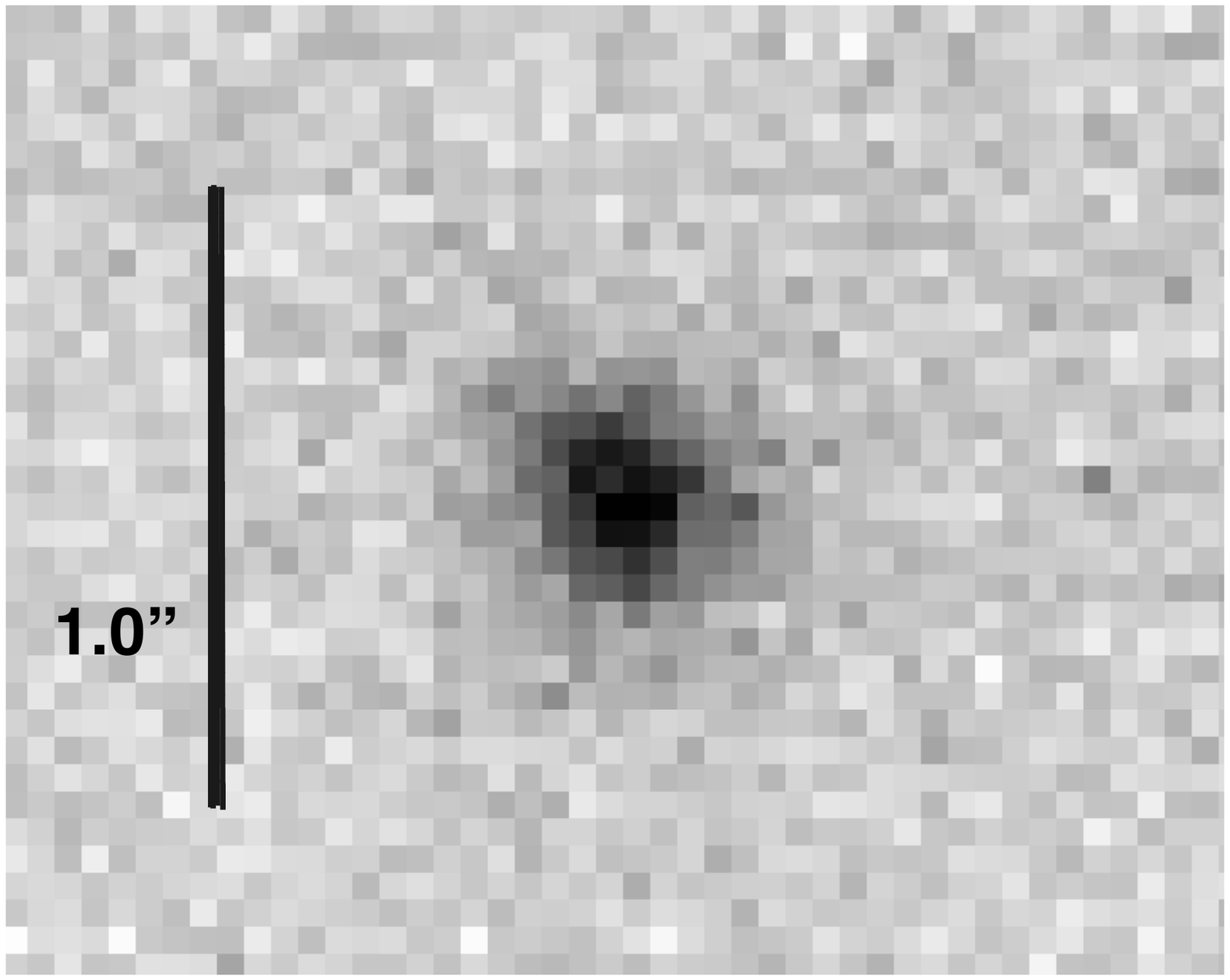}
\vskip -4.4in
\hskip 3.4in
    \includegraphics[angle=0,width=0.475\textwidth]{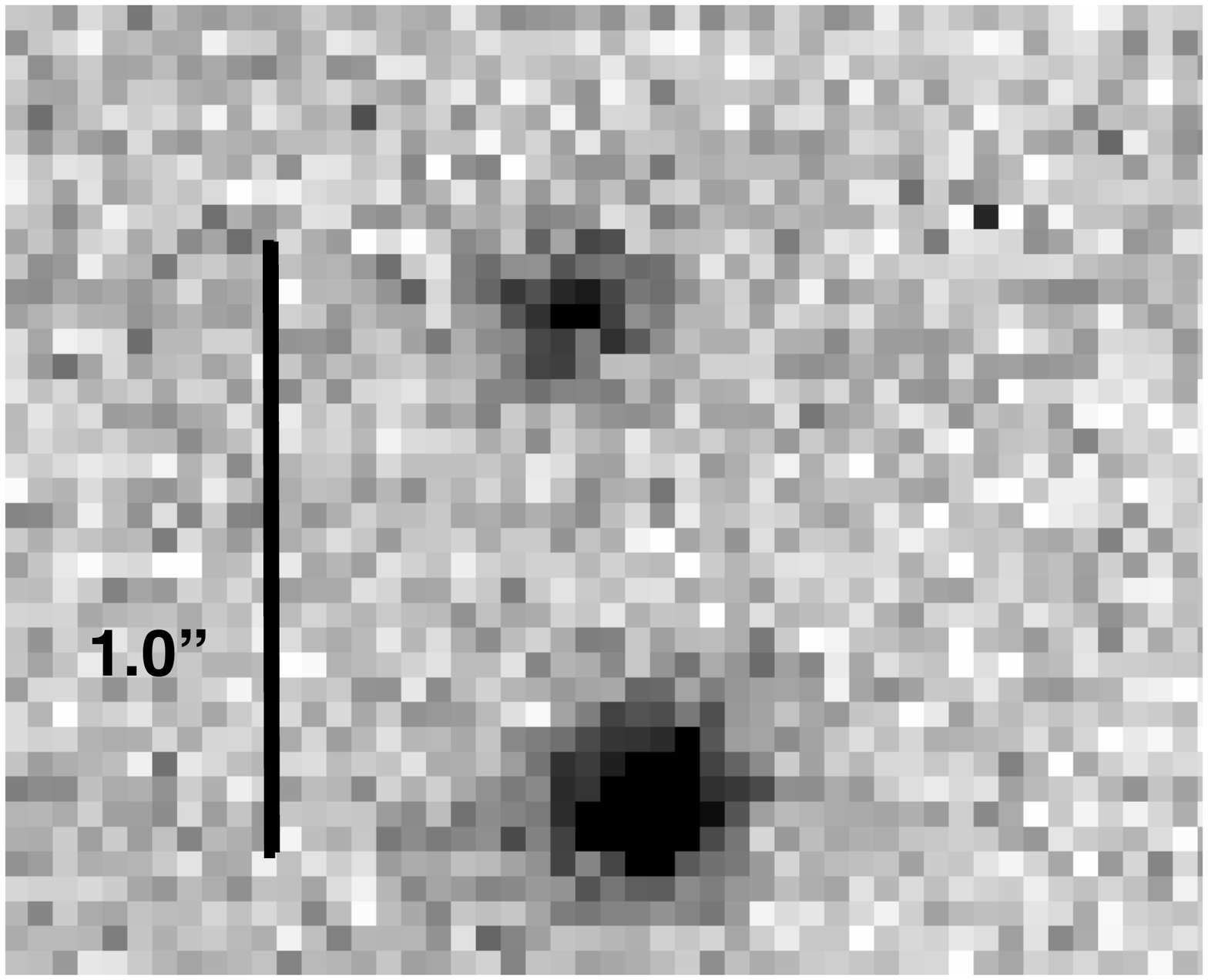}
\vskip -0.8in
\caption{Gemini Observatory $L^{\prime}$ images of 0458AB at 20171224 (left) and 1217AB at 20180102 (right).  North is up and East is left. 
}
\end{figure}

\begin{figure}[!b]
\begin{center}
\vskip -1.0in
\includegraphics[angle=0,width=0.7\textwidth]{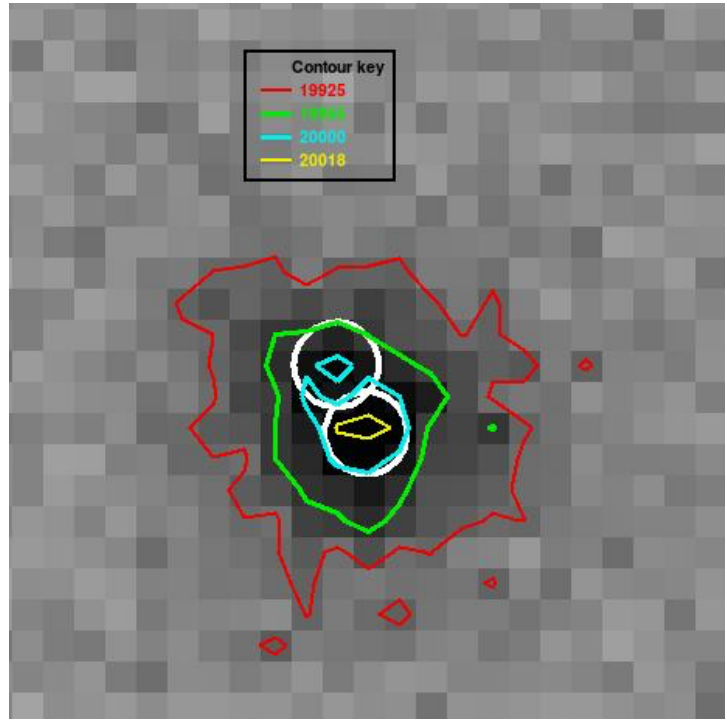}
\vskip -1.2in
\caption{The NIRI $L^{\prime}$  image of 0458AB, with contours overlaid. The white circles are the  $0\farcs12$ diameter apertures used to measure the signal for each component of the binary.
}
\end{center}
\end{figure}

To compensate for the variable sky background at $L^{\prime}$, adjacent frames were subtracted for each target and photometric standard, and then stacked using the known telescope offsets. The single brown dwarfs were observed on nights with seeing typically $1\farcs0$. Aperture photometry with annular sky regions was carried out using apertures of diameter $0\farcs7$ --  $1\farcs0$, corresponding to the seeing on the night of observation.  Aperture corrections were determined from the photometric standards observed close in time and airmass to the science target.

The two binary systems were observed on nights of excellent $0\farcs25$ seeing. Figure 3 shows the $L^{\prime}$ images of the two binary systems. 
The components of 0458AB were found to be very close at the epoch of observation. Figure 4 shows the $L^{\prime}$ image of 0458AB smoothed by
 $2 \times 2$ pixel binning. Contours are overlaid, and the two photometric on-source apertures of  $0\farcs12$ diameter are shown. The contributions to the signal from the background and the other binary component were determined from the radial profile of each source and the outer pixels, for each component. 
1217AB is well-separated in the NIRI images (Figure 3) and aperture photometry with annular skies was carried out, using an aperture diameter of  $0\farcs24$. 
Aperture corrections were determined from the observations of the photometric standards. 
For both binary systems, the magnitude measurement for each component is consistent with a large-aperture measurement of the system, within the uncertainties.

\begin{deluxetable}{lccl}[!t]
\tabletypesize{\small}
\tablewidth{0pt}
\tablecaption{Separation, and Position Angle, of the 1217AB System}
\tablehead{
\colhead{Date UT}  &
\colhead{Separation mas}  &
\colhead{PA degrees} &
\colhead{Instrument/Observatory}
}
\startdata
2012 Jan 29\tablenotemark{a}    &   758.0  $\pm$   0.8     &  14.37  $\pm$    0.07 & NIRC2$+$LGS/Keck II\\
2018 Jan 02   &   881  $\pm$ 3        &  8.7   $\pm$   0.3 & NIRI/Gemini North
\enddata
\tablenotetext{a}{Weighted mean of six measurements in different filters, \citet{Liu2012}.}
\end{deluxetable}

The 0458AB system has shown significant motion compared to previous imaging, and in the next Section we constrain its orbit. The 1217AB system does not show significant motion and the orbit cannot be constrained. For future reference, Table 2 gives the separation and position angle of this binary measured from the NIRI image presented here, and the values of these parameters in 2012, measured by \citet{Liu2012} using Keck laser guide star adaptive optics \citep[LGS AO,][]{Bouchez2004,Wizinowich2006}. Table 3 lists our $L^{\prime}$ photometry for the nine brown dwarfs.

\begin{deluxetable}{llr@{ $\pm$ }l}
\tablewidth{0pt}
\tablecaption{New $L^{\prime}$ Photometry}
\tablehead{
\colhead{Name} & \colhead{Spectral} & \multicolumn{2}{c}{$L^{\prime}$} \\ 
\colhead{} & \colhead{Type} & \multicolumn{2}{c}{mag} 
}
\startdata  
WISE J000517.48$+$373720.5 & T8.5 & 14.43 & 0.10 \\
WISEPA J031325.96$+$780744.2 & T9 & 14.13 & 0.11 \\
WISEA J041022.75$+$150247.9 & Y0 &  15.39 & 0.06 \\
WISEP J045853.90$+$643452.6A  & T8.5 & 14.50  & 0.10 \\
WISEP J045853.90$+$643452.6B  & T9 & 14.81 & 0.15 \\
UGPS J072227.51$-$054031.2\tablenotemark{a}  & T9 & 13.13 & 0.15 \\
WISE J085510.83$-$071442.5\tablenotemark{a} & Y$1+$ & 16.31 & 0.15 \\
WISEPC J121756.91$+$162640.2A & T9 &  14.68 & 0.02 \\
WISEPC J121756.91$+$162640.2B & Y0 & 15.62 & 0.05 \\
WISEA J140518.32$+$553421.3 & Y0.5 & 15.56 & 0.11 \\
WISEA J205628.88$+$145953 & Y0 &  14.87 & 0.12 
\enddata 
\tablenotetext{a}{$L^{\prime}$ photometry was synthesized from flux-calibrated spectra.}
\end{deluxetable}

\subsection{Synthetic $L^{\prime}$ Photometry}

$L^{\prime}$ photometry was synthesized from spectra of  0722 and WISE J085510.83−071442.5 (hereafter 0855). The $2.8 \leq \lambda~\mu$m $\leq 4.2$ spectrum of 0722 was published by \citet{Leggett2012} and 
the $3.40 \leq \lambda~\mu$m $\leq 4.13$ spectrum of 0855 was published by \citet{Morley2018}; each was
flux calibrated using {\it Spitzer} [3.6] photometry. We assumed a zero flux contribution for $3.1 \leq \lambda~\mu$m $\leq 3.4$ for 0855 (see Figure 1). Table 3 lists the synthesized $L^{\prime}$ photometry. Uncertainties were determined from the noise in the spectra and the uncertainty in the calibration photometry.

\section{The Orbit of 0458AB}

\subsection{Astrometric Monitoring of 0458AB}

\subsubsection{Keck LGS AO}

We obtained resolved images of 0458AB at four epochs using the facility infrared imager NIRC2 with the LGS AO system at the Keck~II telescope \citep{Bouchez2004,Wizinowich2006}. 
For the first two epochs in 2011 and 2012 
we used NIRC2's wide camera ($39.686\pm0.008$\,mas\,pix$^{-1}$). The second two epochs in 2018 were obtained after the Gemini $L^{\prime}$   imaging with NIRC2's narrow camera ($9.971\pm0.004$\,mas\,pix$^{-1}$).
For the 2011 and 2018 data sets we used the CH$_4$s filter, centered on the $H$-band flux peak for T~dwarfs ($\lambda = 1.592~\mu$m,
$\Delta\lambda = 0.126~\mu$m), and for the 2012 data set we used the $Y$-band filter (see Appendix of \citealp{Liu2012}).

\begin{figure*}[!b]
\vskip 0.15in
\centerline{
\includegraphics[width=1.5in]{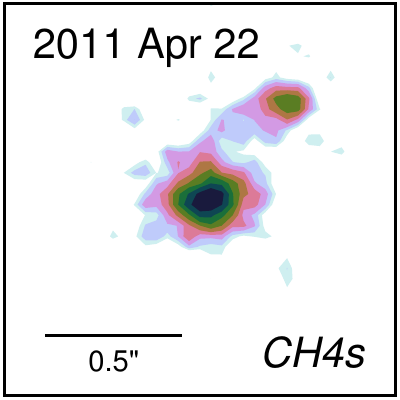}
\includegraphics[width=1.5in]{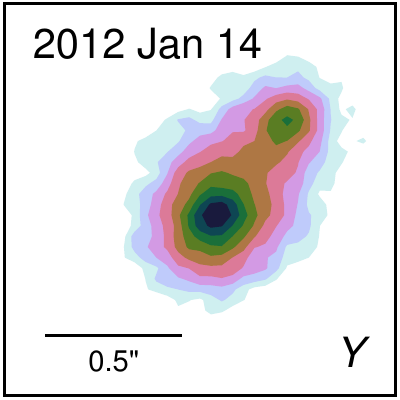}
\includegraphics[width=1.5in]{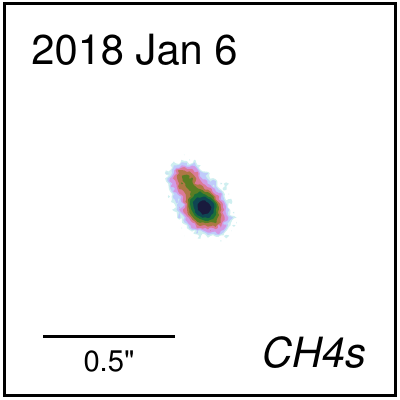}
\includegraphics[width=1.5in]{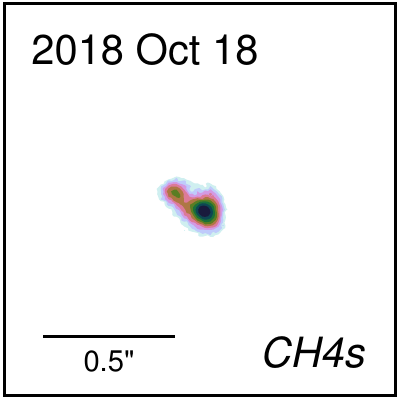}
}
\caption{Contour plots of our Keck LGS~AO images of 0458AB.  Contours are in logarithmic intervals from unity to 10\% of the peak flux.  North is up and East is left.  The separation of the binary changed dramatically, requiring much higher angular resolution imaging at the latest epochs.\label{fig:keck}}
\end{figure*}

We derived binary parameters from our imaging data in the same fashion as in our past work, using a three-component 2D-Gaussian model for PSF fitting and adopting the rms among individual images at a given epoch for the uncertainties in those parameters \citep[e.g.,][]{Dupuy2014,Liu2006}. To convert the instrumental $(x,y)$ measurements into angular separations and position angles (PAs), we used the same methods as in \citet{Dupuy2016} and \citet{Dupuy2017}. We used the calibration of Fu et al.\ (2012, priv.\ comm.)\footnote{\url{http://astro.physics.uiowa.edu/~fu/idl/nirc2wide/}} for the first two  epochs and the calibration of
\citet{Service2016} for the second two  epochs.
We add $-0\fdg262\pm0\fdg020$ to our PAs as a correction for the orientation of NIRC2 
for data obtained after the AO system realignment in 2015~April \citep{Service2016}, and $-0\fdg252\pm0\fdg009$ prior to that \citep{Yelda2010}.  Figure 5 shows typical  images from all four epochs, and Table 4 reports our derived relative astrometry.

\begin{deluxetable}{lcccccc}[!t]
\tabletypesize{\scriptsize}
\tablewidth{0pt}
\tablecaption{Relative Astrometry and Photometry for 0458AB}
\tablehead{
\colhead{Date UT}            &
\colhead{Tel./Inst.}           &
\colhead{Sep. mas}     &
\colhead{PA degree}         &
\colhead{Filter}               &
\colhead{$\Delta{m}$ mag}    &
\colhead{Ref.}                 }
\startdata
2010~Mar~24  & Keck/NIRC2    & $510   \pm20  $     &     $320  \pm1  $         & $J$, $H$          & $0.98\pm0.08$, $1.00\pm0.09$                & G11 \\
2011~Feb~3   & Keck/OSIRIS   & $493   \pm15  $     &     $321.4\pm1.0$         & $H_{\rm bb}$      & \nodata                                     & B12 \\
2011~Apr~22  & Keck/NIRC2    & $465   \pm 7  $\phn &     $322  \pm3  $         & $CH_4s$           & $1.04\pm0.08$                               & *   \\
2011~Aug~29  & Keck/NIRC2    & $455   \pm 4  $\phn &     $322.9\pm0.4$         & $J$, $H$, $K_s$   & $0.98\pm0.01$, $1.02\pm0.01$, $1.06\pm0.03$ & B12 \\
2012~Jan~14  & Keck/NIRC2    & $432   \pm 3  $\phn &     $323.9\pm0.9$         & $Y$               & $0.88\pm0.04$                               & *   \\
2012~Feb~27  & {\it HST}/ACS-WFC  & $435   \pm 9  $\phn &     $324.2\pm0.4$         & $F814W$, $F850LP$ & $0.80\pm0.11$, $0.72\pm0.02$                & *   \\
2015~Jan~2   & {\it HST}/WFC3-IR  & $273.3 \pm 0.7$\phn &     $336.8\pm0.4$         & $F140W$           & $0.98\pm0.03$                               & *   \\
2017~Dec~27  & Gemini-N/NIRI & $103   \pm20  $     & \phn$ 29.0\pm1.0$         & $L^{\prime}$           & $0.31\pm0.18$                                       & *   \\
2018~Jan~6   & Keck/NIRC2    & $130.3 \pm 1.8$\phn & \phn$ 31.9\pm0.7$         & $CH_4s$           & $1.10\pm0.04$                               & *   \\
2018~Oct~18  & Keck/NIRC2    & $132.2 \pm 1.1$\phn & \phn$ 59.2\pm0.2$         & $CH_4s$           & $1.035\pm0.017$                             & *   
\enddata
\tablecomments{*~this work, G11:~\citet{Gelino2011}, B12:~\citet{Burgasser2012}. {\it HST} data from GO-12504 (PI: Liu), GO-13705 (PI: Patience).}
\end{deluxetable}

\begin{deluxetable}{lccccccc}[!t]
\tabletypesize{\scriptsize}
\tablewidth{0pt}
\tablehead{
\multicolumn{2}{c}{Observation Date} &
\colhead{Right Ascension}    &
\colhead{Declination}    &
\colhead{$\sigma_{\rm R.A.}$} &
\colhead{$\sigma_{\rm Decl.}$} &
\colhead{Airmass}    &
\colhead{Seeing}     \\
\colhead{UT}   &
\colhead{MJD}  &
\colhead{deg}  &
\colhead{deg}  &
\colhead{mas}  &
\colhead{mas}  &
\colhead{}  &
\colhead{arcsec}}
\tablecaption{Absolute Astrometry of 0458AB in  CFHT/WIRCam Integrated Light\label{tbl:cfht}}
\startdata
2011~Feb~11 & 55603.2484 & 074.72535576 & $+64.58135589$ &  2.8 &  5.4 & 1.408 & 1.10 \\
2011~Sep~18 & 55822.6268 & 074.72556256 & $+64.58140633$ &  2.8 &  4.0 & 1.415 & 0.77 \\
2011~Sep~23 & 55827.6297 & 074.72556591 & $+64.58140813$ &  2.8 &  2.4 & 1.409 & 0.83 \\
2013~Oct~20 & 56585.5370 & 074.72581431 & $+64.58158644$ &  1.6 &  2.1 & 1.416 & 0.65 \\
2013~Dec~11 & 56637.4207 & 074.72577737 & $+64.58160575$ &  2.2 &  6.1 & 1.409 & 0.97 \\
2014~Oct~9  & 56939.5323 & 074.72594986 & $+64.58166391$ &  2.0 &  3.5 & 1.463 & 0.61 \\
2014~Oct~13 & 56943.5591 & 074.72594815 & $+64.58166721$ &  4.8 &  3.0 & 1.415 & 0.71 \\
2014~Oct~15 & 56945.5697 & 074.72594499 & $+64.58166731$ &  2.9 &  2.5 & 1.409 & 0.64 
\enddata
\tablecomments{The quoted uncertainties correspond to relative, not absolute, astrometric errors.}
\end{deluxetable}

\subsubsection{{\it HST} Imaging}

We analyzed archival {\it HST} images from two epochs. On 2012~Feb~27~UT the program GO-12504 (PI: Liu) observed 0458AB with ACS-WFC as a PSF reference source, and on 2015~Jan~2~UT the program GO-13705 (PI: Patience) obtained pre-imaging with WFC3-IR for their spectroscopic observations. 
For the ACS-WFC data, we use only the higher signal-to-noise ratio (SNR) $F850LP$ imaging. We analyzed the ACS-WFC images as described in Section~3.1.2 of \citet{Dupuy2017}, using TinyTim-based \citep{Krist2011} PSF-fitting.  For the WFC3-IR data, we 
used the appropriate TinyTim PSFs for that instrument as in our 
previous work \citep[e.g.,][]{Dupuy2009a,Dupuy2009b,Liu2008}.  We used the {\tt D2IMARR} and {\tt WCSDVARR} FITS extensions and the {\tt CD} matrices of the headers to convert our measured $(x,y)$ into separations and PAs. Table 4 reports the mean and rms of the mean obtained from individual exposures as our best-fit values and uncertainties.

\subsubsection{CFHT/WIRCam}

We obtained eight epochs of wide-field, unresolved imaging of 0458AB using the facility infrared camera WIRCam
\citep{Puget2004}  at the Canada-France-Hawaii Telescope (CFHT) as part of our ongoing Hawaii Infrared Parallax Program. We used an exposure time of 60\,s in the $J$ band and achieved SNR = 40--70.  We measured $(x,y)$ positions using SExtractor \citep{Bertin1996}  and converted these to 
relative astrometry using a custom pipeline described in our previous work \citep{Dupuy2012,Liu2016}.  The absolute calibration of the linear terms of our astrometric solution was derived by matching low proper motion sources ($<30$\, mas~yr$^{-1}$) 
to the 2MASS point source catalog \citep{Cutri2003}. 
To convert our relative parallax and proper motion to an absolute frame, we use 
the mean parallax and proper motion of stars simulated by the Besan\c{c}on model of the Galaxy \citep{Robin2003}, selecting stars over the same range of apparent magnitudes as in the data. The variance in the conversion from relative to absolute is determined by using many different subsets of model stars.
The resulting astrometry for 0458AB in integrated light is given in Table 5.

\begin{figure}[!b]
\vskip -2.3in
\hskip 0.6in
\includegraphics[angle=0,width=1.05\textwidth]{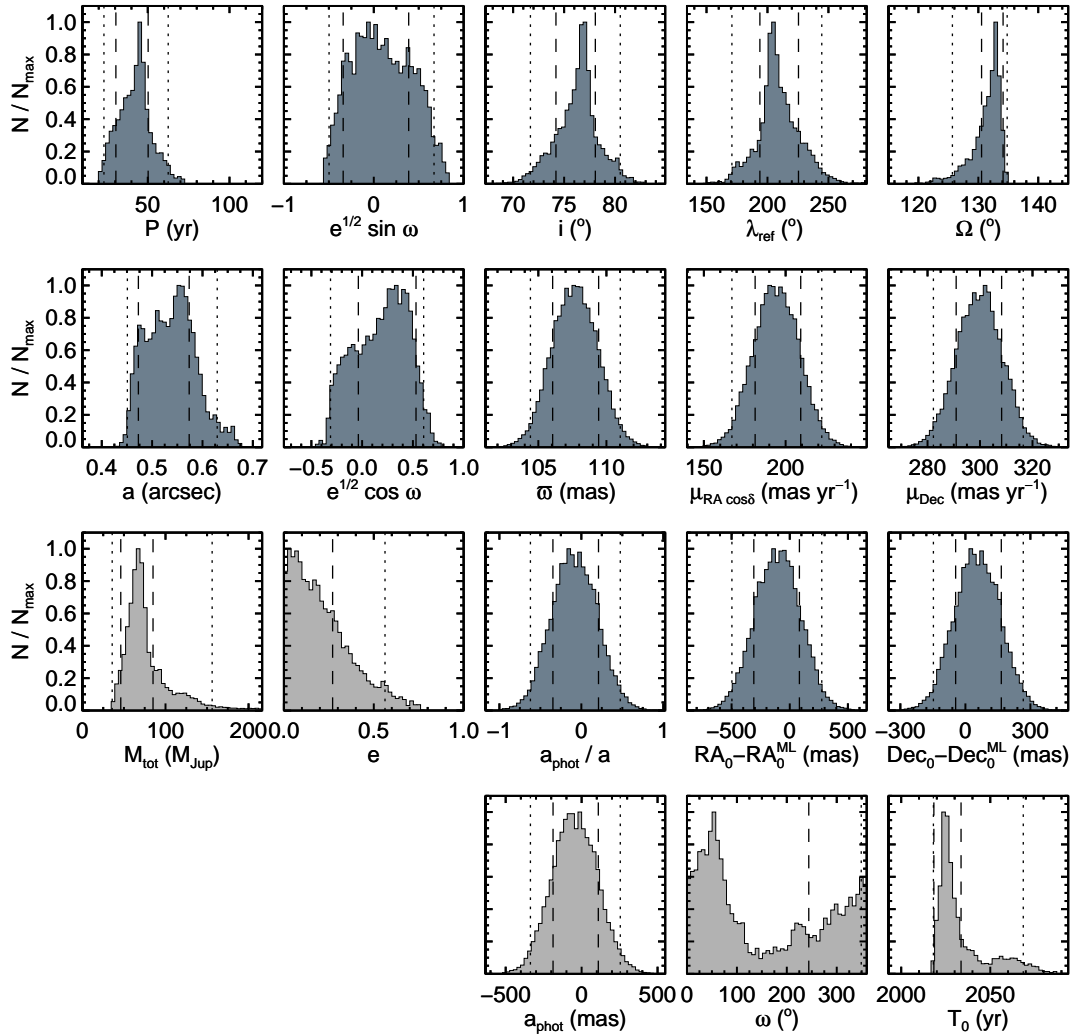}
\vskip -1.9in
\caption{Marginalized posterior distributions for our PT-MCMC analysis of the 0458AB orbit. Dark gray histograms are directly fitted parameters, and light gray histograms are properties computed from the fits.
}
\label{fig:mcmc-hist}
\end{figure}

\begin{deluxetable}{lccc}[!t]
\tablecaption{PT-MCMC Orbital Posteriors for WISE~J0458$+$6434AB \label{tbl:mcmc-WISEPC_J0458+64}}
\setlength{\tabcolsep}{0.10in}
\tabletypesize{\scriptsize}
\tablewidth{0pt}
\tablehead{
\colhead{Property}              &
\colhead{Median $\pm$1$\sigma$} &
\colhead{95.4\% c.i.}           &
\colhead{Prior}                 }
\startdata
\multicolumn{4}{c}{Fitted parameters} \\[1pt]
\cline{1-4}
\multicolumn{4}{c}{} \\[-5pt]
Orbital period, $P$ [yr]                                                     & $43_{-12}^{+7}$                  &           23, 63           & $1/P$ (log-flat)                                                   \\[3pt]
Semimajor axis, $a$ [mas]                                                    & $540_{-70}^{+40}$                &          450, 630          & $1/a$ (log-flat)                                                   \\[3pt]
$\sqrt{e}\sin{\omega}$                                                       & $0.1_{-0.4}^{+0.3}$              &       $-$0.5, 0.7          & uniform                                                            \\[3pt]
$\sqrt{e}\cos{\omega}$                                                       & $0.22_{-0.25}^{+0.31}$           &      $-$0.31, 0.61         & uniform                                                            \\[3pt]
Inclination, $i$ [\degree]                                                   & $76.5_{-2.3}^{+1.5}$             &         71.7, 80.5         & $\sin(i)$, $0\degree < i < 180$[\degree]                             \\[3pt]
PA of the ascending node, $\Omega$ [\degree]                               & $132.0_{-1.5}^{+2.1}$            &        125.7, 134.8        & uniform                                                            \\[3pt]
Mean longitude at $t_{\rm ref}=2455197.5$~JD, $\lambda_{\rm ref}$ [\degree]  & $207_{-14}^{+18}$                &          171, 245          & uniform                                                            \\[3pt]
${\rm R.A.}_{\rm ref}-{\rm R.A.}_{\rm ref}^{\rm ML}$ [mas]                               & $-40_{-90}^{+80}$                &       $-$210, 120          & uniform, ${\rm R.A.}_{\rm ref}^{\rm ML} =  74.7254631$             \\[3pt]
${\rm decl.}_{\rm ref}-{\rm decl.}_{\rm ref}^{\rm ML}$ [mas]                             & $50_{-100}^{+110}$               &       $-$150, 270          & uniform, ${\rm decl.}_{\rm ref}^{\rm ML} = +64.5813263$            \\[3pt]
Relative proper motion in R.A., $\mu_{\rm R.A., rel}$ [mas~yr$^{-1}$]               & $194_{-13}^{+15}$                &          167, 222          & uniform                                                            \\[3pt]
Relative proper motion in decl., $\mu_{\rm decl., rel}$  [mas~yr$^{-1}$]             & $300_{-9}^{+8}$                  &          282, 316          & uniform                                                            \\[3pt]
Relative parallax, $\varpi_{\rm rel}$ [mas]                                  & $107.7_{-1.6}^{+1.8}$            &        104.4, 111.0        & $1/\varpi^2$                                                       \\[3pt]
Ratio of photocenter orbit to semimajor axis, $a_{\rm phot}/a$               & $-0.09_{-0.26}^{+0.30}$          &      $-$0.62, 0.47         & uniform                                                            \\[3pt]
\cline{1-4}
\multicolumn{4}{c}{Computed properties} \\[1pt]
\cline{1-4}
\multicolumn{4}{c}{} \\[-5pt]
Eccentricity, $e$                                                            & $0.18_{-0.18}^{+0.09}$           &         0.00, 0.56         & \nodata                                                            \\[3pt]
Argument of periastron, $\omega$ [\degree]                                   & $110_{-110}^{+130}$              &            0, 350          & \nodata                                                            \\[3pt]
Time of periastron, $T_0=t_{\rm ref}-P\frac{\lambda-\omega}{360\degree}$ [JD]& $1994_{-18}^{+12}$               &         1964, 2018         & \nodata                                                            \\[3pt]
Photocenter semimajor axis, $a_{\rm phot}$ [mas]                             & $-50_{-140}^{+160}$              &       $-$340, 250          & \nodata                                                            \\[3pt]
$(a^3 P^{-2})\times10^4$ [arcsec$^3$ yr$^{-2}$]                              & $0.85_{-0.29}^{+0.16}$           &         0.44, 1.87         & \nodata                                                            \\[3pt]
Correction to absolute R.A. proper motion, $\Delta\mu_{\rm R.A.}$  [mas~yr$^{-1}$]   & $0.48_{-0.13}^{+0.14}$           &         0.19, 0.76         & \nodata                                                            \\[3pt]
Correction to absolute decl. proper motion, $\Delta\mu_{\rm decl.}$ [mas~yr$^{-1}$]  & $-0.76\pm0.18$                   &      $-$1.13, $-$0.43      & \nodata                                                            \\[3pt]
Correction to absolute parallax, $\Delta\varpi$ [mas]                        & $0.583_{-0.022}^{+0.023}$        &        0.541, 0.630        & \nodata                                                            \\[3pt]
Absolute proper motion in R.A., $\mu_{\rm R.A.}$\tablenotemark{a}  [mas~yr$^{-1}$]                    & $195_{-13}^{+15}$                &          168, 223          & \nodata                                                            \\[3pt]
Absolute proper motion in decl., $\mu_{\rm decl.}$\tablenotemark{a}  [mas~yr$^{-1}$]                 & $299\pm9$                        &          281, 315          & \nodata                                                            \\[3pt]
Absolute parallax, $\varpi$\tablenotemark{a} [mas]                                            & $108.3_{-1.6}^{+1.7}$            &        105.1, 111.7        & \nodata                                                            \\[3pt]
Distance, $d$ [pc]                                                           & $9.24_{-0.15}^{+0.14}$           &         8.96, 9.52         & \nodata                                                            \\[3pt]
Semimajor axis, $a$ [AU]                                                     & $5.0_{-0.6}^{+0.3}$              &          4.2, 5.9          & \nodata                                                            \\[3pt]
Total mass, $M_{total}$ [$M_{\rm Jup}$]                                                  & $70_{-24}^{+15}$                 &           36, 156          & \nodata                                                           
\enddata
\tablenotetext{a}{The absolute parallax, proper motion in RA and proper motion in Declination are consistent with the values determined from {\it Spitzer} images by \citet{Kirkpatrick2019}, which are 
$109.2\pm 3.6$, $207.7 \pm 1.2$ and	$291.2 \pm 1.2$, respectively.
\tablecomments{The full 13-parameter fit has $\chi^2 = 28.1$ (23 dof), and the relative orbit has $\chi^2 = 17.5$ (13 dof). The orbit quality metrics defined by \citet{Dupuy2017} are $\delta\log M_{total} = 0.26$\,dex, $\delta{e} = 0.27$, and $\Delta{t_{\rm obs}/P} = 0.20$, indicating a poorly constrained orbit determination. {\tt ML} is Maximum Likelihood.}
}
\end{deluxetable}

\begin{deluxetable}{lcclcc}[!t]
\tabletypesize{\small}
\tablewidth{0pt}
\tablecaption{Predicted Separation, and Position Angle, of the 0458AB System}
\tablehead{
\colhead{Date UT}  & \colhead{Separation mas}  & \colhead{PA degrees} & \colhead{Date UT}  & \colhead{Separation mas}  & \colhead{PA degrees} 
}
\startdata
2019 Sep 1 & $160.8\pm 1.7$ & $ 83.0\pm 0.5$ & 2022 Jan 1 & $286  \pm18  $ & $112.1\pm 2.6$ \\
2020 Jan 1 & $176.9\pm 2.3$ & $ 89.6\pm 0.6$ & 2022 Mar 1 & $294  \pm21  $ & $113.3\pm 3.1$ \\
2020 Mar 1 & $185.3\pm 2.7$ & $ 92.4\pm 0.7$ & 2022 Sep 1 & $318  \pm34  $ & $117  \pm 6  $ \\
2020 Sep 1 & $213  \pm 4  $ & $ 99.7\pm 1.0$ & 2023 Jan 1 & $330  \pm40  $ & $119  \pm12  $ \\
2021 Jan 1 & $231  \pm 6  $ & $103.5\pm 1.2$ & 2023 Mar 1 & $340  \pm50  $ & $120  \pm14  $ \\
2021 Mar 1 & $240  \pm 7  $ & $105.2\pm 1.3$ & 2023 Sep 1 & $360  \pm60  $ & $123  \pm19  $ \\
2021 Sep 1 & $268  \pm13  $ & $109.6\pm 1.8$ & 2024 Jan 1 & $370  \pm70  $ & $126  \pm22  $ \\
           &                &                & 2024 Mar 1 & $380  \pm70  $ & $127  \pm23  $  
\enddata
\end{deluxetable}

\begin{figure}[!b]
\vskip -5.3in
\hskip 0.33in
\includegraphics[angle=0,width=1.0\textwidth]{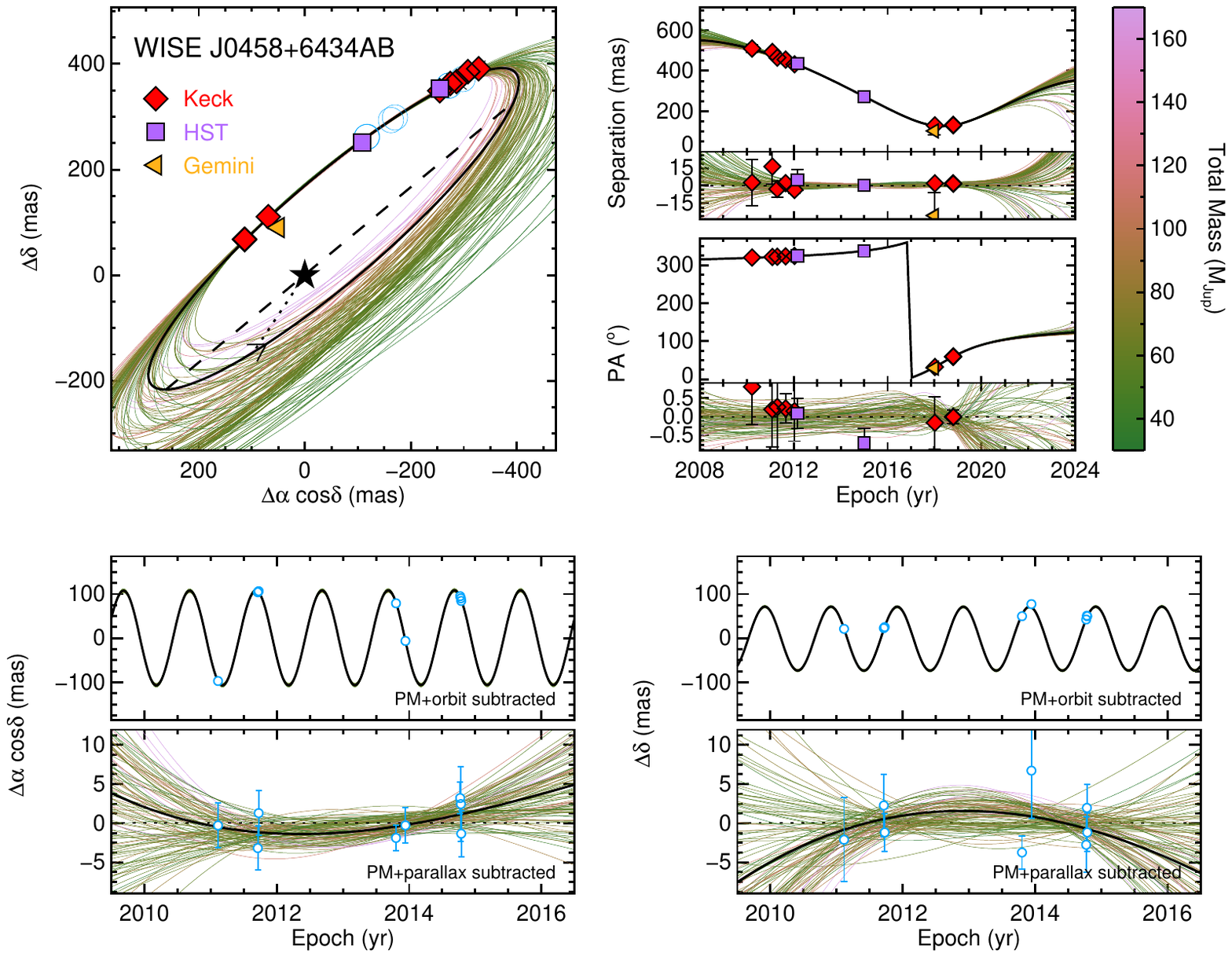}
\vskip -0.05in
\caption{Orbital analysis for 0458AB. The highest-likelihood orbit is a thick black line, and 100 randomly drawn PT-MCMC solutions are thin lines color-coded by total mass. {\bf Top left:} Relative astrometry from Keck LGS AO  (red diamonds), {\it HST} (purple squares), and Gemini (gold triangle). Open circles mark the times corresponding to CFHT/WIRCam observations. The dashed line is the line of nodes, and the arrow indicating motion direction is plotted at periastron. {\bf Top right:} Relative astrometry as a function of time with the lower subpanels showing residuals from the highest-likelihood orbit. {\bf Bottom:} Integrated-light astrometry from CFHT/WIRCam as a function of time. Upper subpanels show the parallax curve after subtracting proper motion and orbital motion (error bars are too small to be visible). Lower subpanels show the orbital motion after subtracting proper motion and parallax. This is for display purposes only, as our analysis fits proper motion, parallax, and orbital motion simultaneously. \label{fig:orbit}}
\end{figure}

\subsection{The Orbit, Parallax, and Proper Motion of 0458AB}

We combined our resolved astrometry with other published measurements and our integrated-light astrometry  in a single analysis, fitting the orbit, parallax, and proper motion. The approach is very similar to our past work \citep{Dupuy2015,Dupuy2017}. Six of the thirteen parameters are shared between the resolved and integrated-light data, all relating to orbit: period ($P$), eccentricity ($e$) and  argument of periastron ($\omega$) parametrized as $\sqrt{e}\sin{\omega}$ and $\sqrt{e}\cos{\omega}$, inclination ($i$), PA of the ascending node ($\Omega$), and mean longitude at the reference epoch ($\lambda_{\rm ref}$), defined to be 2010~January~1~00:00~UT (2455197.5~JD). There are two parameters for orbit size; the semimajor axis ($a$) in angular units, and the ratio of the semimajor axis of the CFHT photocenter orbit to $a$ ($a_{\rm phot}/a$). The five remaining parameters are all related to the CFHT astrometry: parallax ($\varpi_{\rm rel}$), proper motion ($\mu$) in Right Ascension and Declination, and the Right Ascension and Declination at the reference epoch $t_{\rm ref}$. The only parameters without uniform priors were $P$ and $a$ (log-flat), $i$ ($\sin{i}$, random viewing angles), and an approximately uniform space density ($\varpi_{\rm rel}^{-2}$).

We use the parallel-tempering Markov chain Monte Carlo (PT-MCMC) ensemble sampler in \texttt{emcee~v2.1.0}
\citep{Foreman2013} that is based on the \citet{Earl2005} algorithm. ``Hot'' chains explore essentially
all of the allowed parameter space between solutions, while ``cold'' chains find local minima. Information is exchanged
between chains and the solution is the  ``coldest'' of 30 chains. We use 100 walkers to sample our 13-parameter model over $8\times10^4$ steps. The initial state is a random, uniform draw over all of parameter space for bounded parameters: $e$, $\omega$, $\Omega$, $i$, $\lambda_{\rm ref}$; $2 < P/{\rm yr} < 2000$; $0\farcs01 < a < 1\farcs0$; $-1 < a_{\rm phot}/a < 1$; $\pm$100\,mas around 
the reference epoch Right Ascension and Declination; $\pm$30\% around the 
relative proper motion; and $\pm$20\% around the 
relative parallax.  The resulting 
distributions of posteriors  are shown in Figure 6 and summarized in Table 6. Figure 7 displays the orbit and Table 7 gives estimates of future configurations of the system.

{\it It should be noted that the total dynamical mass is not well measured from the current data spanning 8.6 years. The derived mass is dependent on the choice of priors for parameters such as period, semimajor axis, and eccentricity.} Under our current assumptions, the minimum system mass is 36~$M_{\rm Jup}$ (2$\sigma$), suggesting that neither of the components is planetary mass ($< 13 M_{\rm Jup}$). One reliable prediction from our orbit analysis is that the separation will continue increasing for the next few years, at least until 2021, the nominal launch year of {\it JWST} (Figure 7, Table 7). Our analysis also provides the first parallax measurement for 0458AB that properly accounts for orbital motion, although we do not detect significant astrometric perturbations in our CFHT data over more than three years.


\section{Constraints on the Properties of 0458AB and 1217AB from Photometry}

\begin{figure}[!b]
\begin{center}
\vskip -0.2in
\includegraphics[angle=0,width=0.7\textwidth]{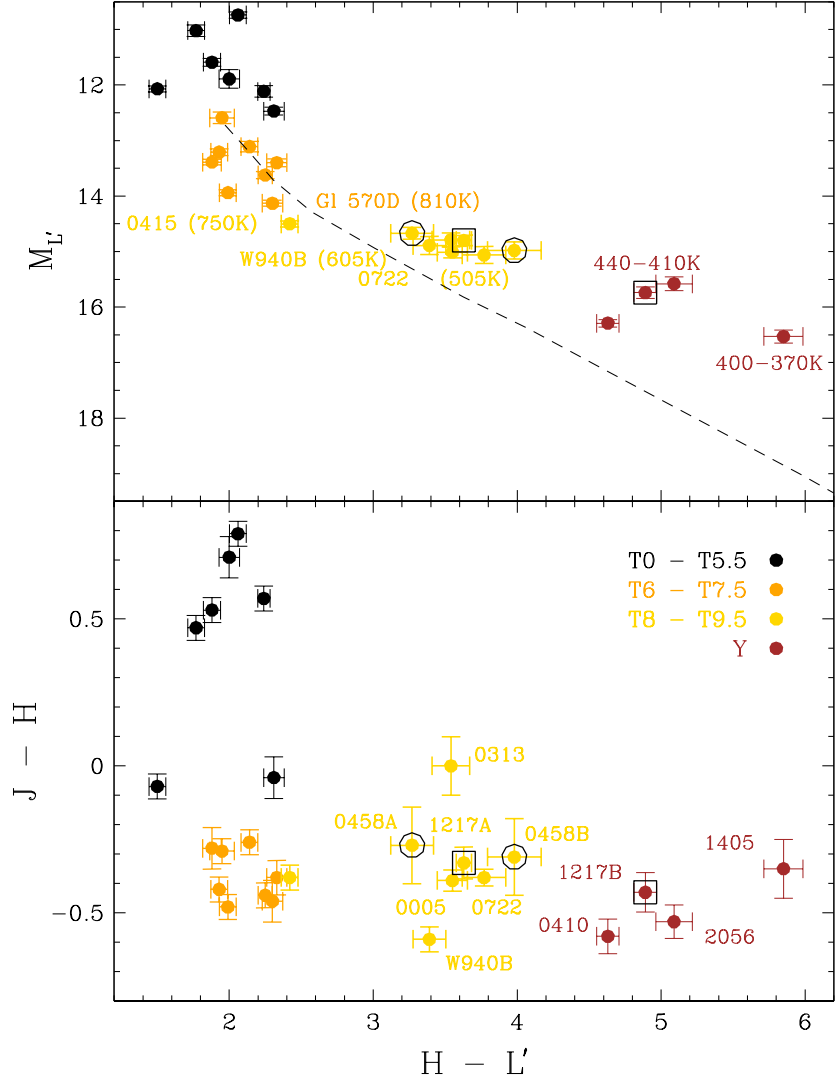}
\vskip -0.35in
\caption{$JHL^{\prime}$ colors of T and Y dwarfs.  Open circles indicate the components of the 0458AB system, and open squares indicate the components of the 1217AB system. In the upper plot, well-studied T dwarfs are identified and their $T_{\rm eff}$ values are given (Section 4). The $T_{\rm eff}$ ranges for the Y dwarfs are also shown \citep[derived from near-infrared spectra and mid-infrared photometry,][]{Leggett2017}. The reddest (coolest) objects are identified in the lower plot. 
The dashed line in the upper plot is a model sequence from \citet[][see Section 5]{Tremblin2015}. 
}
\end{center}
\end{figure}

Figure 8 is a color-magnitude and color-color plot using the 1 -- 4~$\mu$m  photometry of T and Y dwarfs. Brown dwarfs with $T_{\rm eff}$ $<$ 750~K ($H - L^{\prime} > 2.5$)
show a  brightening at $4~\mu$m relative to the near-infrared.
The components of the 0458AB and 1217AB binaries have colors typical of late-T and Y dwarfs, suggesting that their composition and age are typical of a field population. 
Figure 8 shows that, as would be expected, the T8.5 0458A lies in a similar region of the color-magnitude diagrams as the T8.5  Wolf 940B, and the T9 0458B lies in a similar region as the T9 0722. The T9 1217A also lies in a similar region as the T9 0722, and the Y0 1217B is similar to the Y0 2056.

\begin{deluxetable}{lccccccc}[!t]
\tabletypesize{\small}
\tablecaption{Physical Properties of the 0458AB and 1217AB Systems\tablenotemark{a}}
\tablehead{
\colhead{} & \multicolumn{7}{c}{$M_{\rm Jup}$, log $g$} \\
\colhead{Name} & \colhead{Age=0.6} & \colhead{1.0} & \colhead{3.0} & \colhead{4.0} & \colhead{6.0} & \colhead{8.0} & \colhead{10.0~Gyr} 
}
\startdata
0458A & 9, 4.4 & 15, 4.5 & 25, 4.8 & 28, 4.9 & 34, 5.0 & 37, 5.1 & 41, 5.2 \\ 
0458B  & 12, 4.3 & 11, 4.4 & 19, 4.7 & 22, 4.8 & 26, 4.9 & 29, 5.0 & 32, 5.0 \\
1217A  & 12, 4.3 & 11, 4.4 & 19, 4.7 & 22, 4.8 & 26, 4.9 & 29, 5.0 & 32, 5.0 \\
1217B & 7, 4.1 & 9, 4.3 & 15, 4.6 & 17, 4.6 & 20, 4.7 & 23, 4.8 & 25, 4.9 
\enddata 
\tablenotetext{a}{If 0458A has $T_{\rm eff} \approx 600$~K, 0458B and 1217A have  $T_{\rm eff} \approx 500$~K, and 1217B has  $T_{\rm eff} \approx 425$~K; using  \citet{Saumon2008} evolutionary models.}
\end{deluxetable}

Wolf 940B is a benchmark object, with age and composition constrained by its distant M dwarf companion. A large amount of data is available for Wolf 940B, including a mid-infrared spectrum from {\it Spitzer}. The studies by \citet{Burningham2009} and \citet{Leggett2010b} show this T8.5 to have $T_{\rm eff} = 605 \pm 20$~K, log $g = 5.0 \pm 0.2$  (cm s$^{-2}$) and a metallicity within 0.2~dex of solar. 0722 is bright and has also been well studied, although it does not have a stellar companion or a spectrum beyond $4~\mu$m. \citet{Leggett2012} and \citet{Lucas2010} find for 0722 that $T_{\rm eff} = 505 \pm 10$~K,  log $g = 4.0 \pm 0.5$  and metallicity is close to solar. For the Y dwarfs, the study of near-infrared spectra and mid-infrared photometry by \citet{Leggett2017} shows that 2056 has $T_{\rm eff} = 425 \pm 15$~K,  log $g = 4.5 \pm 0.25$ and metallicity is solar  or slightly super-solar. 

Using Wolf 940B, 0722 and 2056 as reference objects, the photometric comparison indicates that 0458A has $T_{\rm eff} \approx 600$~K, 0458B and 1217A have  $T_{\rm eff} \approx 500$~K, and 1217B has  $T_{\rm eff} \approx 425$~K; the values for the 1217AB system are consistent with previous analyses \citep[e.g.,][]{Leggett2014}. Assuming that the binary components have the same age, evolutionary models can be used to constrain gravities and masses for the two systems. Table 8 lists these values as a function of age. 

Our preliminary orbit for 0458AB gives a total mass for the system of
$70^{+15}_{-24}$~$M_{\rm Jup}$ ($1 \sigma$, Table 6).  Combining the astrometry with the observed photometric difference of $\Delta(J) = 0.98 \pm 0.01$ mag \citep{Burgasser2012} 
constrains the individual masses to 
$57_{-28}^{+25}$\,$M_{\rm Jup}$ and
$14_{-22}^{+21}$\,$M_{\rm Jup}$ ($1 \sigma$).
Note that when the photocenter orbit $a_{phot}$ is poorly constrained, as it is here (Table 6), the uncertainties in the individual masses are large.
Assuming coevality and $T_{\rm eff} \approx 600$~K for the primary and $\approx 500$~K for the secondary, the evolutionary models give a broad age range for this system of 3 -- 13~Gyr (Table 8).  For the 1217AB system, fits to the near-infrared spectrum and mid-infrared photometry  of 1217B 
constrain the likely age to be 0.7 -- 6~Gyr \citep{Leggett2017}.  The tangential velocities of the 0458AB and 1217AB systems are $16\pm 1$~km~s$^{-1}$ (Table 6) and $62 \pm 6$~km~s$^{-1}$ \citep{Leggett2017} respectively, suggesting thin disk membership and an age $<$ 10~Gyr \citep{Dupuy2012,Robin2003}. Adopting a likely age of a few Gyr for both systems, the masses of the primary and secondary are around 35 and 25~$M_{\rm Jup}$ for 0458AB, and around  20 and 15~$M_{\rm Jup}$ for 1217AB.

\section{The $\lambda\approx 4$~micron Problem}

All available atmospheric models predict fluxes in the $4~\mu$m region that are too low for the late-T and Y dwarfs. These include the cloud-free non-equilibrium chemistry models of \citet{Marley2002}, the cloud-free chemical equilibrium models with updated opacities of \citet{Saumon2012},   the cloudy  chemical equilibrium models of \citet{Morley2012,Morley2014a}, and the cloud-free  non-equilibrium chemistry models with updated opacities of \citet{Tremblin2015}, as demonstrated by 
\citet{Leggett2010a, Leggett2012, Leggett2013, Leggett2015, Leggett2017} and \citet{Luhman2016}.
The discrepancy is illustrated in Figure 1 for the $T_{\rm eff} = 500$~K object 0722 \citep{Leggett2012}. 

In Figure 1 we show the observed spectrum, and synthetic spectra generated by the models of \citet{Morley2012} and \citet{Tremblin2015}, for this 500~K brown dwarf.  
These two sets of model grids are the best available at this temperature, at the time of writing; the former includes clouds but not the non-equilibrium chemistry brought about by mixing, and the latter includes non-equilibrium chemistry but does not include clouds (work on a grid of model atmospheres that includes both clouds and non-equilibrium chemistry is ongoing by members of our team \citep{Marley2017}).
\citet{Morley2012,Morley2014a} show that clouds of chloride and sulfide condensates are important for $400 \lesssim T_{\rm eff}$~K $\lesssim 900$, and  water clouds are important for $T_{\rm eff} \lesssim 300$~K. The effect of the clouds is primarily a reduction in the $\lambda \sim 1~\mu$m flux with that energy redistributed to longer wavelengths \citep{Morley2012,Morley2014a}. 
Vertical mixing in brown dwarf atmospheres leads to an increase in the abundances of the more stable CO and N$_2$, and a decrease in the abundances of the less stable CH$_4$ and NH$_3$ 
\citep[e.g.,][]{Saumon2006}. The resulting non-equilibrium chemistry has been shown to be important for both T and Y dwarfs \citep{Leggett2007,Leggett2015,Saumon2006,Saumon2007,Stephens2009}. The decrease in NH$_3$ absorption leads to 
an increase in near-infrared flux, especially in the $H$-band, and an increase in flux at $\lambda \sim 10.5~\mu$m, while the increase in CO absorption leads to less flux at $\lambda \sim 4.5~\mu$m
\citep[e.g. Figure 1;][]{Saumon2006,Morley2014a}. 

Figure 1 suggests that cloudy models are required to reproduce the $Y$-band flux, which  is the wavelength most impacted by clouds at this temperature \citep{Morley2014a}. Non-equilibrium chemistry is required to reproduce the $H$-band shape and the 4.5~$\mu$m flux. Neither model reproduces the shape of the $K$-band flux peak and both models are deficient at $\lambda \sim 4~\mu$m. In this work we use as a primary reference the non-equilibrium models of \citet{Tremblin2015}; this is because the dominant opacities at $\lambda\sim 4~\mu$m consist of
carbon- and nitrogen-bearing molecules (Figure 1), and clouds (as currently modelled) do not significantly impact this wavelength region. 

The upper panel of Figure 8 explores the $H - L^{\prime}$ colors of T and Y dwarfs, and shows the
color sequence generated by the \citet{Tremblin2015}  models. This comparison suggests that 
the $\lambda \sim 4~\mu$m flux discrepancy starts at $T_{\rm eff} \approx 700$~K and increases to lower temperatures. Figure 8 suggests that at $L^{\prime}$ the models are too faint by $\sim 0.6$~mag at 500~K, and too faint by $\sim 1.6$~mag at 400~K.

\begin{figure}[!b]
\begin{center}
\vskip -0.2in
\includegraphics[angle=0,width=0.7\textwidth]{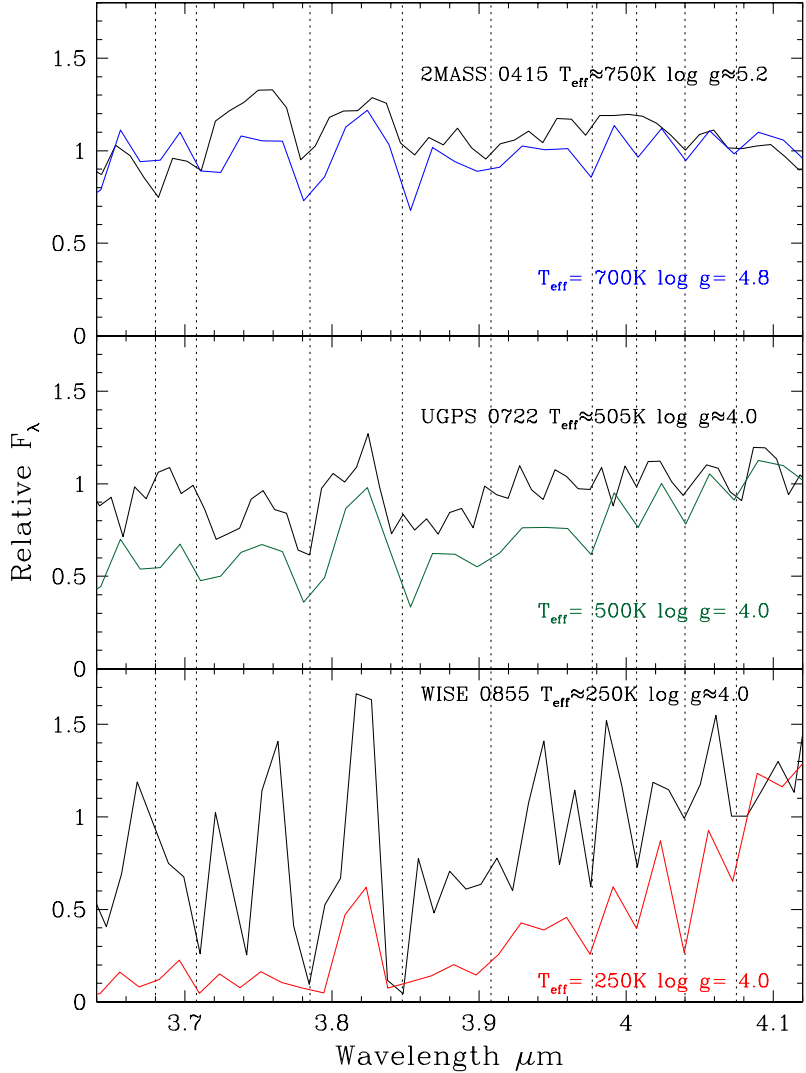}
\vskip -0.4in
\caption{Observed spectra (black lines) and calculated spectra (colored lines) at $\lambda \sim 3.9~\mu$m for brown dwarfs with $T_{\rm eff} \sim$ 250, 500 and 750~K.  To approximately flatten the spectra, they  have been divided by a cubic function derived from the 0722 spectrum. The spectra were scaled so that $F_{\lambda} \approx 1.0$ at $\lambda \approx 4.08~\mu$m. Dotted vertical lines indicate CH$_4$ absorption features
\citep{Tennyson2012,Yurchenko2014}. 
Although the CH$_4$ features 
map well between the observations and the models, the flux between the absorption bands
is much lower in the models --- for example at $\lambda \approx$ 3.69, 3.75, 3.82, 3.88 and 3.94~$\mu$m
the 500~K model flux is 
too low by a factor of $\sim 1.5$ and the 250~K model flux is too low by factors of 3 -- 10.
}
\end{center}
\end{figure}

To explore this further, Figure 9  compares observed $L$-band spectra to synthetic spectra generated by \citet{Tremblin2015} models.  Observed spectra are shown for 2MASS J04151954$-$0935066 \citep[hereafter 0415,][]{Sorahana2012}, 0722 \citep{Leggett2012}, and 0855 \citep{Morley2018}. The synthetic spectra have atmospheric parameters similar to those of the three targets (see the Figure 9 legends). 
The top panel of Figure 9 show that, as expected, the synthetic and observed spectra agree quite well at $T_{\rm eff} \sim 750$~K.  However for cooler atmospheres there is a significant discrepancy. Although the principal opacity appears to be CH$_4$ in both the observed and synthetic spectra, the observed slope is flatter than the calculated slope.  The strong absorption features at $3.7 \leq \lambda~\mu$m $\leq 3.9$ (see also Figure 1)  deepen with decreasing $T_{\rm eff}$, but the fluxes between the absorption features are much higher than calculated.

In their analysis of the cold brown dwarf 0855, \citet{Morley2018} find that the 3.5 -- 4.1~$\mu$m and  4.5 -- 5.1~$\mu$m spectra
can be fit by metal-poor models with a C/O ratio half solar (although the models are then too bright in the near-infrared). 
As pointed out by \citet{Morley2018}, it is unlikely that all the late-T and Y dwarfs have such an unusual atmospheric composition and so it is more likely that there is something occurring in these cool atmospheres that is not captured by the models. \citet{Leggett2017} explored changes to the adiabatic index in models, such that the deep atmosphere was warmer and the upper atmosphere was cooler than the standard model. These experiments could improve the agreement with observations in the $K$-band ($\lambda = 2.1~\mu$m, see Figure 1) but the discrepancy at [3.6] remained. 

\begin{figure}[!b]
\begin{center}
\includegraphics[angle=0,width=0.95\textwidth]{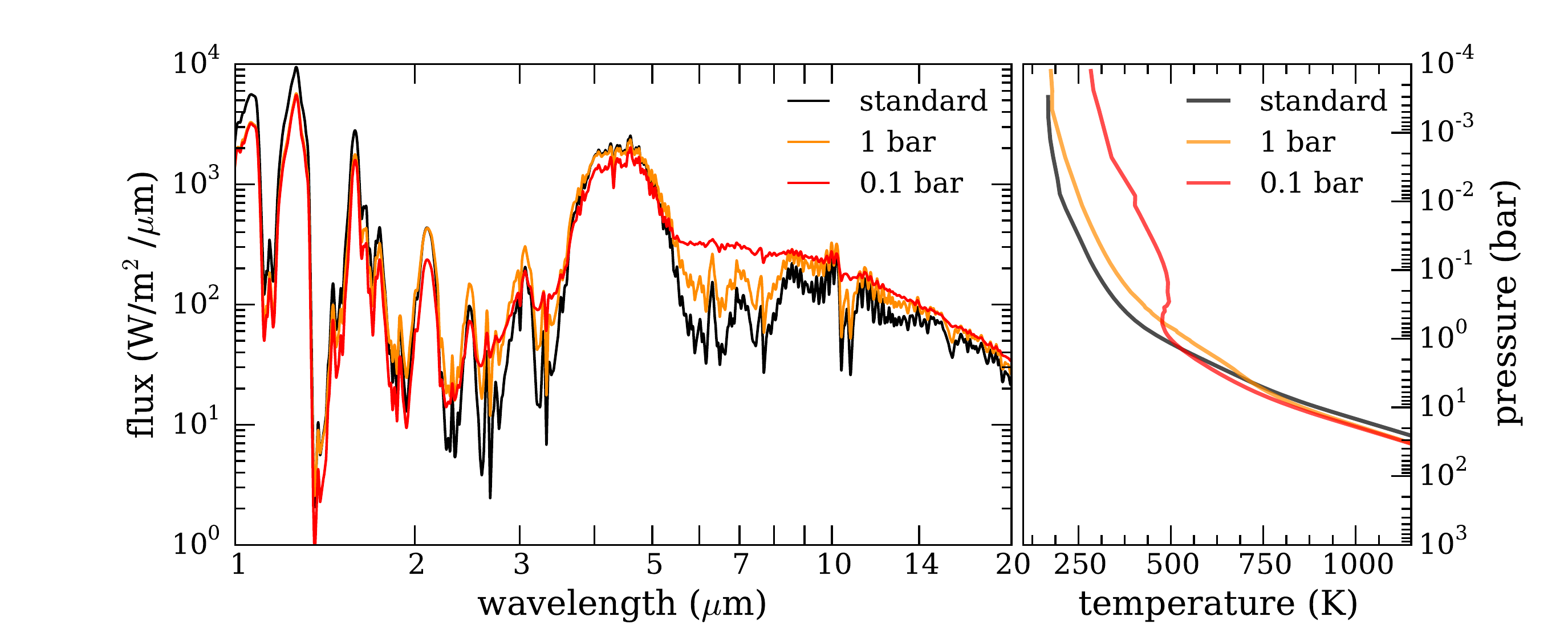}
\vskip -0.1in
\caption{A synthetic 500~K spectrum (left) and Pressure-Temperature profile (right) demonstrating the effect of adding heat at 1 or 0.1 bar by adding energy at those altitudes over a scale height. 
}
\end{center}
\end{figure}

The fact that the observed flux is higher than calculated suggests that the   3.6 -- 4.1~$\mu$m flux is emerging from warmer atmospheric layers than the models generate.
Figure 10 shows a $T_{\rm eff} = 500$~K synthetic spectrum generated by us, based on the models of \citet{Morley2014a},  which demonstrates the changes in the spectrum that could be brought about if the atmosphere is heated at 1 bar or 0.1 bar. The pressure-temperature profile for this atmosphere is also shown, illustrating the size of the temperature differential in the upper atmosphere. The heated-atmosphere spectrum is very similar to the standard spectrum in the near-infrared and at $\lambda \approx 5~\mu$m, but is much brighter at $\lambda \approx 3.5~\mu$m and at $\lambda \approx 6$ -- $8~\mu$m.  This preliminary result suggests that upper atmosphere heating in late-T and Y dwarfs could be the cause of the brighter than expected W1, [3.6] and $L^{\prime}$ magnitudes, and may also give rise to the (less well-defined) discrepancy seen in the W3 
magnitudes  \citep[$7.5 \lesssim \lambda~\mu$m $\lesssim 16.5$,][Figure 12]{Leggett2017}. The total emitted flux  increases, and $T_{\rm eff}$ increases from 500~K to 550~K; this implies that temperatures determined for late-T and Y dwarfs by fitting models to  red, near-infrared and  4.5~$\mu$m data could be significantly too low. 
However, this heated-atmosphere model is exploratory only, and needs further study.

A cool brown dwarf atmosphere is assumed to be 
undergoing adiabatic cooling in the deep atmosphere and radiative cooling in the upper atmosphere \citep[e.g.,][Figure 1]{Marley2015}.
 The retrieval analysis of late-T and Y dwarf atmospheres by \citet{Zalesky2019} found temperature structures largely consistent with radiative-convective equilibrium, and chemical abundances for water, methane and ammonia to be as expected. Their analysis explored fits to spectra covering $0.9 \lesssim \lambda~\mu$m $\lesssim 1.7$ only, which is sensitive to deep atmospheric layers and not very sensitive to the upper atmosphere \citep[Figure 10 and][Figure 2]{Zalesky2019}. Also, as the authors state, the derived gravities are uncomfortably high and radii uncomfortably low, suggesting that sampling this limited wavelength range is not providing reliable parameters.

Energy (or heat) could be introduced into a brown dwarf atmosphere by thermochemical instabilities 
\citep{Tremblin2015,Tremblin2019} or cloud clearing \citep{Morley2012}. Interestingly, 
measurements of the atmospheres of 
the solar system giant planets show the upper layers to be warmer than expected; heat sources such as breaking gravity waves have been invoked \citep{Matcheva1999,ODonoghue2016}. The same effect may be present in cold brown dwarfs, which have similar radii and rotation periods \citep{Cushing2016,Leggett2016,Manjavacas2019},
and highly dynamic atmospheres \citep{Apai2017,Showman2013}.

It is also important to note that these models are one-dimensional, and it is likely that the atmospheres have both horizontal and vertical pressure/temperature variations. Variability at $1 \lesssim \lambda~\mu$m $\lesssim 5$ has been measured at the few-percent level for T and Y0 dwarfs \citep[e.g.,][]{Buenzli2012,Buenzli2014,Cushing2016,Leggett2016,Manjavacas2019,Metchev2015}. The variability may be due to irregular cloud cover \citep[e.g.,][]{Marley2010,Morley2014b}, or it may be due to the presence of zones and spots, similar to the solar system giant planets \citep[e.g.,][]{Apai2017,Tan2017}. A one-dimensional model fit over the entire wavelength range may underpredict the flux at wavelengths where isolated hot spots are bright.

\section{Conclusion}

We have imaged two brown dwarf binary systems at high angular resolution using NIRI and its $L^{\prime}$ filter on the Gemini North telescope:  the T8.5 $+$ T9 0458AB, and the T9 $+$ Y0 1217AB. We have also imaged five single brown dwarfs in  $L^{\prime}$ at lower angular resolution: 0005 (T8.5), 0313 (T9), 0410 (Y0), 1405 (Y0.5), and 2056 (Y0). In addition, we have synthesized  $L^{\prime}$ photometry from published spectra for 0722 (T9) and 0855 (Y1$+$). 

The 0458AB system has shown significant orbital motion. The separation of the components  was  $0\farcs$46 in 2011 and  
$0\farcs$13 in 2018, a decrease in projected separation from 4.3~AU to 1.2~AU. We have combined the Gemini images with higher resolution Keck LGS AO and {\it HST} images  
to monitor the orbit of the 0458AB system, and with wide-field CFHT images to determine the proper motion and parallax of the system.  Our preliminary orbital analysis gives a period of $43_{-12}^{+7}$ years and a total mass for the system of $70_{-24}^{+15} ~M_{\rm Jup}$ ($1 \sigma$). Our analysis will aid the acquisition of the target for ${\it JWST}$ observations. The orbital analysis, together with photometry and evolutionary models, suggests that the age of the system is a few Gyr with component masses of around 35 and 25~$M_{\rm Jup}$.

We verify  that model fluxes 
at $3.4 \lesssim \lambda~\mu$m $\lesssim 4.1$ are too low, as has been found previously. The discrepancy starts at $T_{\rm eff} \approx 700$~K and gets worse to lower temperatures --- at $T_{\rm eff} = 500$~K model fluxes are about a factor of two too low and at  $T_{\rm eff} = 400$~K
the fluxes are  about a factor of four too low. The spectra suggest that the dominant opacity source in this region is CH$_4$ as expected, and the depths of the features are approximately correct; however, the flux emerging between the features, the pseudo-continuum, is brighter than calculated by the models. We have generated model spectra where heat is introduced into the  upper layers of the atmosphere. Such models can significantly increase the flux at $\lambda \sim 3~\mu$m and 
 $\lambda \sim 7~\mu$m without impacting the near-infrared or  $\lambda \sim 5~\mu$m flux, offering the potential of a much better match to observations.
Departures from pure radiative-convective equilibrium temperature-pressure profiles, such as in the test model, can arise from several physical mechanisms.
\citet{Tremblin2019} use hydrodynamic simulations to show that  a diabatic profile is appropriate in the event of instabilities brought about by the conversion between CO and CH$_4$ for warmer brown dwarfs with $T_{\rm eff} \approx 1000$~K; at   $T_{\rm eff} \approx 500$~K similar instabilities could be introduced by the conversion between N$_2$ and NH$_3$ \citep[e.g.,][Figure 2]{Lodders1999}. Another possible source of heat in the upper atmosphere is breaking gravity waves, as has been proposed to explain the higher than expected temperatures in the upper atmospheres of the solar system giant planets 
\citep[e.g.,][]{Matcheva1999,ODonoghue2016}. 
Three-dimensional hydrodynamic models may be necessary to better understand these atmospheres; 
although computationally challenging, schemes are being developed to make the calculations more tractable \citep[e.g.,][]{Venot2019}.

It is important to resolve the discrepancies between models and observations  at   
$\lambda\sim 3.8~\mu$m, for brown dwarfs with $T_{\rm eff} < 700$~K.
For example, the heated-atmosphere model predicts that the  $6 \lesssim \lambda~\mu$m $\lesssim 8$ flux contributes significantly to the bolometric luminosity (Figure 10), and therefore current estimates of $T_{\rm eff}$ are systematically and significantly low.
We eagerly await   {\it JWST} spectra covering these wavelengths, and anticipate that the $\lambda > 3~\mu$m spectra 
delivered by {\it JWST} and {\it SPHEREx} will reveal unexpected climate physics for cool brown dwarfs. 
This physics is likely to be important not only for the brown dwarfs, but also for exoplanets and the solar system giant planets.

\acknowledgements

This publication makes use of data from the Wide-field Infrared Survey Explorer, a joint project of the University of California, Los Angeles, and the Jet Propulsion Laboratory/California Institute of Technology, funded by the National Aeronautics and Space Administration. This work is based in part on archival data obtained with the Spitzer Space Telescope, operated by the Jet Propulsion Laboratory, California Institute of Technology under a contract with NASA. This work is also based in part on observations made with the NASA/ESA Hubble Space Telescope, obtained from the data archive at the Space Telescope Science Institute. STScI is operated by the Association of Universities for Research in Astronomy, Inc. under NASA contract NAS 5-26555.

Some of the data presented herein were obtained at the W. M. Keck Observatory, which is operated as a scientific partnership among the California Institute of Technology, the University of California and the National Aeronautics and Space Administration. The Observatory was made possible by the generous financial support of the W. M. Keck Foundation. Some of the data presented herein were obtained  with WIRCam, a joint project of CFHT, the Academia Sinica Institute of Astronomy and Astrophysics (ASIAA) in Taiwan, the Korea Astronomy and Space Science Institute (KASI) in Korea, Canada, France, and the Canada-France-Hawaii Telescope (CFHT) which is operated by the National Research Council (NRC) of Canada, the Institut National des Sciences de l'Univers of the Centre National de la Recherche Scientifique of France, and the University of Hawaii.

This work is based on observations obtained at the Gemini Observatory, which is operated by the Association of Universities for Research in Astronomy, Inc., under a cooperative agreement with the NSF on behalf of the Gemini partnership: the National Science Foundation (United States), National Research Council (Canada), CONICYT (Chile), Ministerio de Ciencia, Tecnolog\'{i}a e Innovaci\'{o}n Productiva (Argentina), Minist\'{e}rio da Ci\^{e}ncia, Tecnologia e Inova\c{c}\~{a}o (Brazil), and Korea Astronomy and Space Science Institute (Republic of Korea).

MCL and WMJB acknowledge support from NSF grant AST-1518339.

\clearpage
\bibliography{ms_August12}
\bibliographystyle{aasjournal}

\appendix

In this Appendix we describe the relationship between ground-based $L^{\prime}$ and $M^{\prime}$ photometry and the space-based {\it WISE} 
W1 and W2 and {\it Spitzer} [3.6] and [4.5] photometry, so that datasets can be better utilized. See Figure 1 for the location of the filters with respect to a 500~K brown dwarf spectrum. The {\it Spitzer} data 
are taken from  \citet{Kirkpatrick2019,Leggett2017,Martin2018} and references therein, the trigonometric parallaxes are taken from \citet{Kirkpatrick2019,Leggett2017,Martin2018,Smart2018,Theissen2018} and references therein. In addition, for this work, we determined W1 magnitudes from images downloaded from the unWISE database\footnote{\url{http://unwise.me/imgsearch///}} \citep{Schlafly2019} for four Y dwarfs, 
WISE J033605.05-014350.4 with $W1=18.20 \pm 0.10$, WISEA J035000.31-565830.5 with $W1=18.3 \pm 0.2$,  WISE J064723.23-623235.5 with $W1=18.8 \pm 0.2$, and WISEA J235402.79+024014.1 with $W1=18.1 \pm 0.3$.

\begin{figure}[!b]
\begin{center}
\vskip -0.4in
\includegraphics[angle=-90,width=1.0\textwidth]{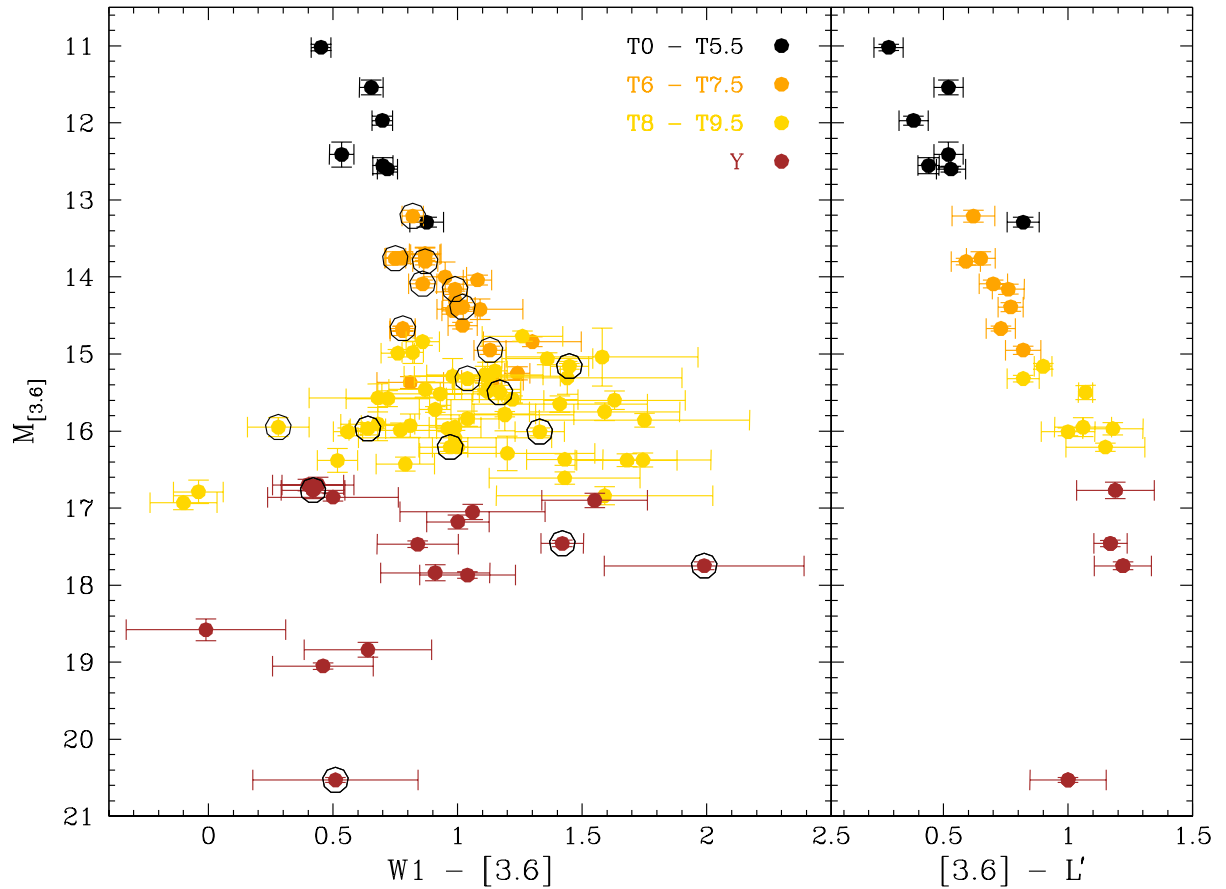}
\vskip -0.35in
\caption{Color-magnitude diagrams for the $4~\mu$m filters.  Circled points in the left panel indicate late-T and Y dwarfs with $L^{\prime}$ photometry that appear in the right panel. 
}
\end{center}
\end{figure}

Figures 11 and 12 are color-magnitude diagrams for the $4~\mu$m and $5~\mu$m filters. We find that the uncertainty in the W1 colors of the T and Y dwarfs may be underestimated, as the large scatter in the W1 $-$ [3.6] color is not seen in the [3.6] $- L^{\prime}$  color. A spot check of the outliers suggests that the W1 photometry is contaminated by nearby background sources --- the brown dwarfs are faint in W1 and the {\it WISE} pixels are large \citep[see also][]{Kirkpatrick2019}. 

The  [3.6] $- L^{\prime}$ color becomes redder for later spectral types, however W2 $-$ [4.5] and $M^{\prime} -$ [4.5] stay close to zero. 
This is not surprising given the large degree of overlap in the W2, [4.5] and $M^{\prime}$ filters (Figure 1).

\begin{figure}[!b]
\begin{center}
\vskip -0.3in
\includegraphics[angle=-90,width=1.0\textwidth]{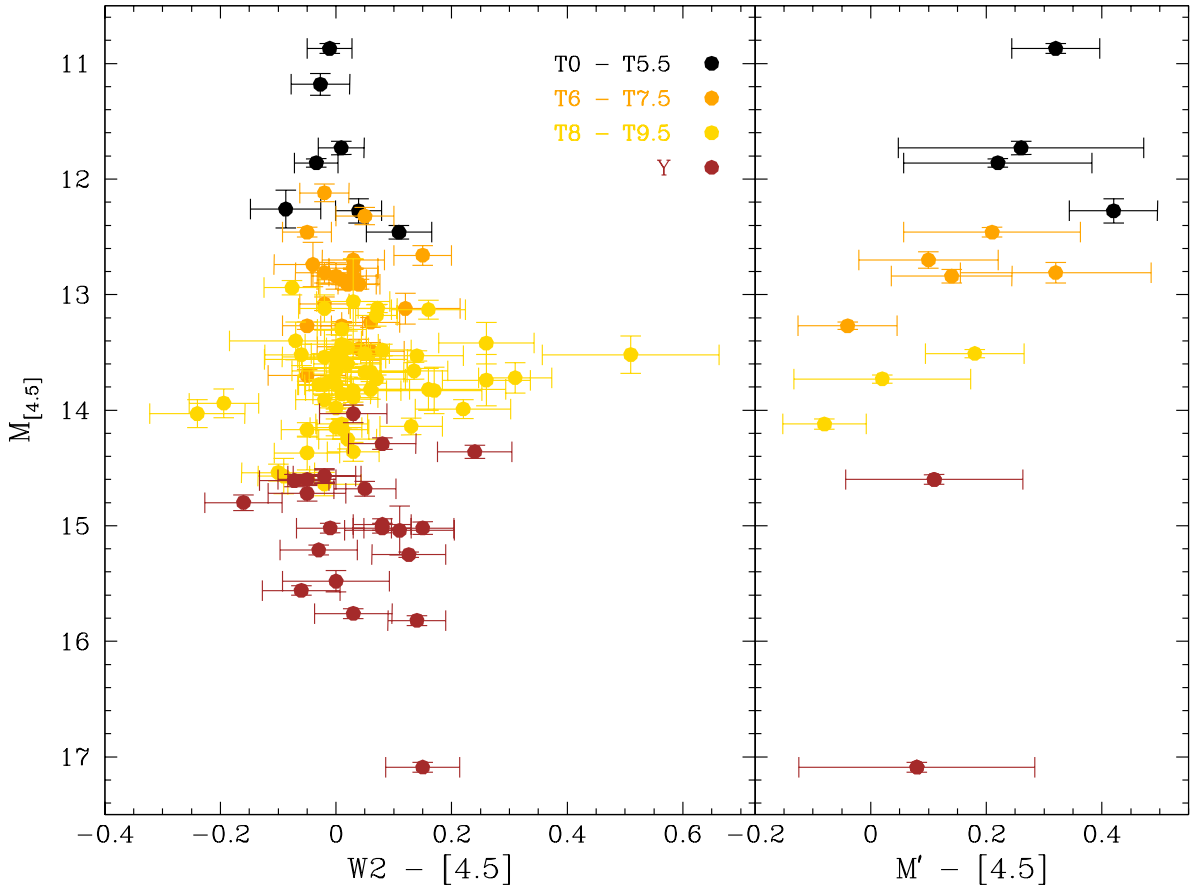}
\vskip -0.35in
\caption{Color-magnitude diagrams for  the $5~\mu$m filters.  Note that the $x$-axis range is much smaller than that of Figure 11 --- these filters give similar magnitudes for the late-T and Y dwarfs.
}
\end{center}
\end{figure}

Figure 13 shows [3.6] $- L^{\prime}$, $L^{\prime} -$ [4.5], $M^{\prime} -$ [4.5], and [3.6] $- M^{\prime}$ as a function of [3.6] $-$ [4.5]. 
Known binaries have been excluded from the sample.
Excluding the extremely red 0855, and the low gravity SDSS J111009.99$+$011613.0 ￼\citep{Gagne2015} which appears discrepant, we find that weighted linear fits can be used to estimate the differences between the ground-based $L^{\prime}$ and  $M^{\prime}$ magnitudes and the {\it Spitzer} [3.6] and [4.5] magnitudes as a function of the [3.6] $-$ [4.5] color:
\vskip 0.1in
\begin{center}
[3.6] $- L^{\prime} = 0.338 + 0.3260\times$([3.6]$-$[4.5])
\end{center}
and
\begin{center}
$L^{\prime} -$ [4.5] $= -0.344 + 0.6919\times$([3.6]$-$[4.5])
\end{center}
for $0.1 \leq$ ([3.6] $-$ [4.5]) $\leq 2.8$. 
\begin{center}
$M^{\prime} -$ [4.5] $= 0.377 - 0.1785\times$([3.6]$-$[4.5])
\end{center}
and
\begin{center}
[3.6] $- M^{\prime} = -0.435 + 1.2163\times$([3.6]$-$[4.5])
\end{center}
for $0.2 \leq$ ([3.6] $-$ [4.5]) $\leq 2.2$.
The rms uncertainty in the linear fit is 0.09~mag for all colors.

\begin{figure}[!b]
\vskip -0.25in
\includegraphics[angle=-90,width=1.0\textwidth]{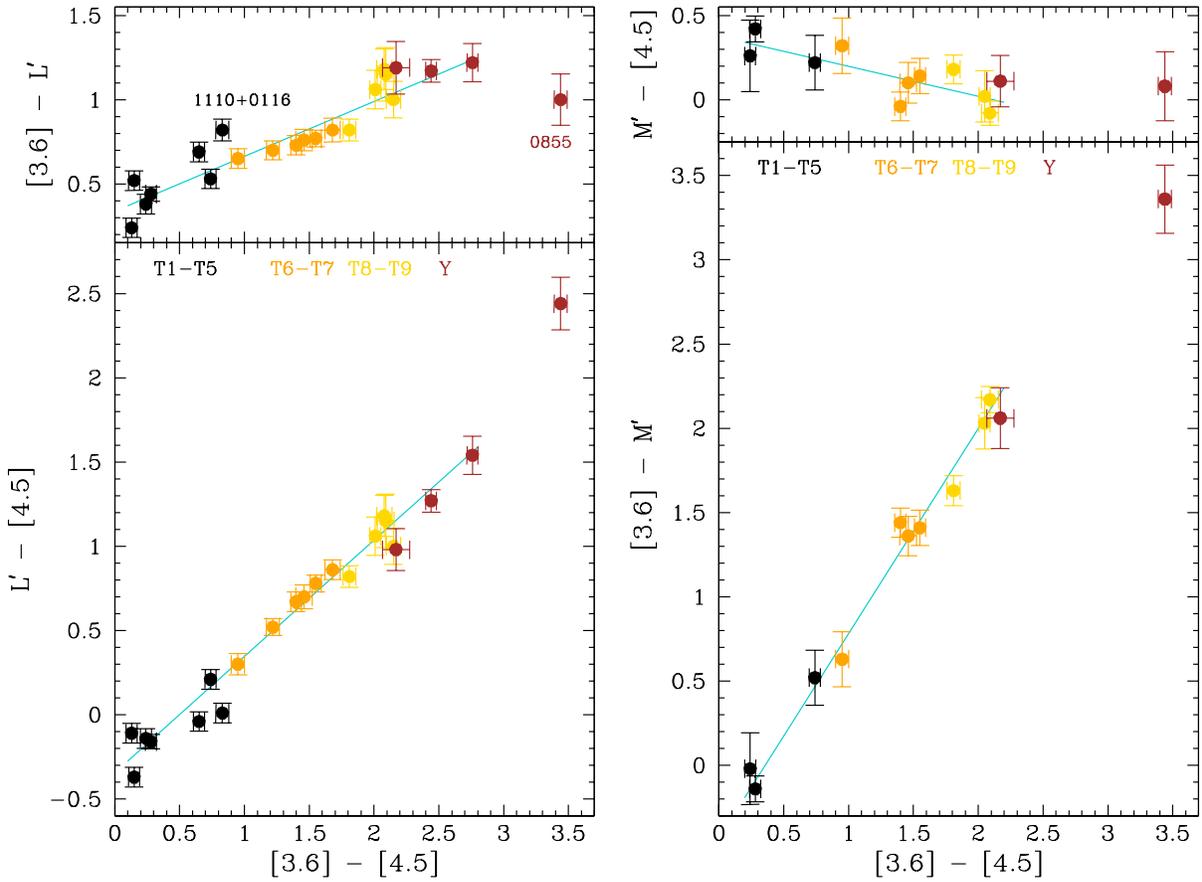}
\vskip -0.4in
\caption{Color-color diagrams for  the 4 and $5~\mu$m filters.  The weighted linear fits shown in cyan exclude SDSS J111009.99$+$011613.0 for the $L^{\prime}$ colors and 0855 for all colors.
}
\end{figure}

Figure 14 shows [3.6] $- L^{\prime}$, $L^{\prime} -$ [4.5], $M^{\prime} -$ [4.5], and [3.6] $- M^{\prime}$
as a function of $H - L^{\prime}$ and $H - M^{\prime}$. Weighted quadratic fits were made, excluding
known binaries, the extremely red 0855, and the low gravity SDSS J111009.99$+$011613.0, as above. 
These relationships can be used to estimate   [3.6] and [4.5] if only $L^{\prime}$  or $M^{\prime}$ are available, for example in the case of close binaries unresolved by  {\it Spitzer}. We use the $H$ bandpass to provide the near-infrared color --- more $H$ measurements are available than $K$ (which can be faint for late-type dwarfs), and shorter wavelengths are impacted by clouds (e.g. Section 5).  We find:
\vskip 0.1in
\begin{center}
[3.6] $- L^{\prime} = -0.255 + 0.5198\times (H - L^{\prime}) - 0.04508\times (H - L^{\prime})^2$
\end{center}
for $1.8 \leq (H - L^{\prime}) \leq 6$, with rms uncertainty 0.08~mag, and
\begin{center}
$L^{\prime} -$ [4.5] $= -0.080 + 0.4081\times (H - L^{\prime}) - 0.02748\times (H - L^{\prime})^2$
\end{center}
for $1.8 \leq (H - L^{\prime}) \leq 6$ and spectral type T6 and later, with rms uncertainty 0.11~mag.
\begin{center}
$M^{\prime} -$ [4.5] $= 0.791- 0.3477\times (H - M^{\prime}) + 0.03769\times (H - M^{\prime})^2$
\end{center}
for $1.0 \leq (H - M^{\prime}) \leq 6$, with rms uncertainty 0.08~mag, and
\begin{center}
[3.6] $- M^{\prime} = -2.290 + 1.7934\times (H - M^{\prime}) - 0.17982\times (H - M^{\prime})^2$
\end{center}
for $1.0 \leq (H - M^{\prime}) \leq 6$, with rms uncertainty 0.13~mag.

\begin{figure}[!b]
\vskip -2in
\hskip -0.3in
\includegraphics[angle=0,width=1.0\textwidth]{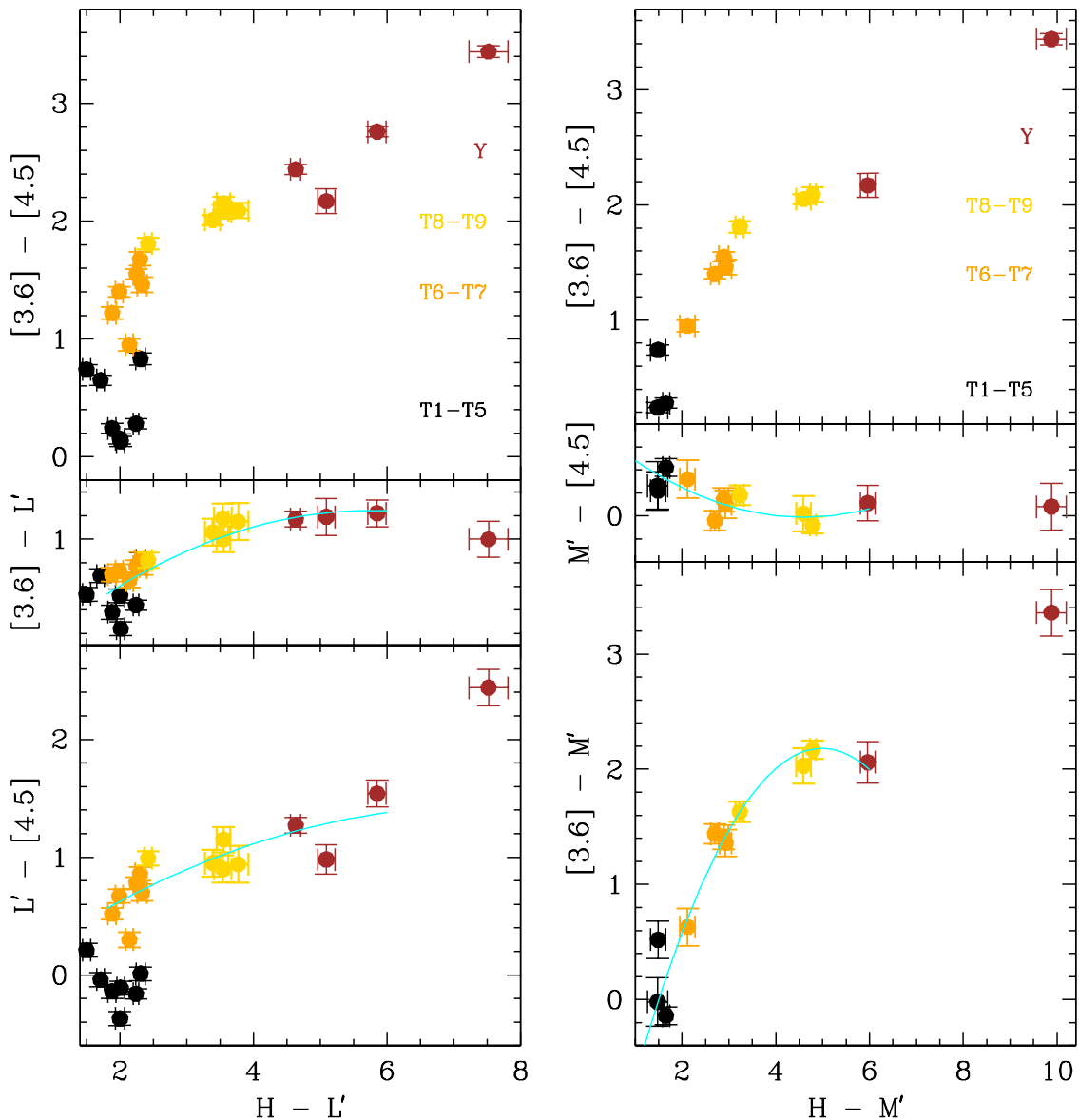}
\vskip -0.8in
\caption{Color-color diagrams for  the $H$, $4~\mu$m and $5~\mu$m filters.  The weighted quadratic fits shown in cyan exclude SDSS J111009.99$+$011613.0  and 0855. Also, the fit to $H - L^{\prime}$:$L^{\prime} -$ [4.5] excludes objects earlier then T6 spectral type due to the rapid increase in $L^{\prime} -$ [4.5] color at $H - L^{\prime} \approx 2$ (see also Figure 8).
}
\end{figure}

\end{document}